\documentclass[nofootinbib,amsmath,prd,aps,superscriptaddress,tightenlines,12pt]{revtex4-2}

\usepackage{graphicx}% Include figure files
\usepackage{dcolumn}% Align table columns on decimal point
\usepackage{bm}% mathbf math
\usepackage{epsfig}
\usepackage{amsmath}
\input{epsf}
\usepackage{psfrag}
\usepackage{xcolor}
\usepackage{ulem}
\usepackage{subcaption}
\newcommand{\bea}{\begin{eqnarray}}
\newcommand{\eea}{\end{eqnarray}}
\newcommand{\nn}{\nonumber}

\begin{document}

\title{The one-point charge correlator in deep inelastic scattering}

\author{Haotian Cao}
\email{haotiao.cao@northwestern.edu}
 \affiliation{Department of Physics \& Astronomy, Northwestern University, Evanston, Illinois 60208, USA}
\affiliation{Joint BNL-SBU Center for Frontiers in Nuclear Science (CFNS),
Stony Brook University, Stony Brook, New York 11794, USA}
\author{Frank~Petriello}
\email{f-petriello@northwestern.edu}
 \affiliation{Department of Physics \& Astronomy, Northwestern University, Evanston, Illinois 60208, USA}

\begin{abstract}

In this work, we propose a novel definition of the one-point charge correlator (QC) adapted to the Breit frame in deep-inelastic scattering (DIS). We demonstrate that this observable is infrared and collinear (IRC) safe, ensuring its perturbative calculability. Utilizing soft-collinear effective theory (SCET), we systematically analyze the QC in both the forward and back-to-back limits. In the forward limit, we introduce the nucleon charge correlator as a novel non-perturbative object that encodes the multi-dimensional microscopic structure of the nucleon. In the back-to-back limit, the QC establishes a direct correspondence with transverse momentum-dependent distributions (TMDs), enabling its description within the standard TMD factorization formalism. The singular distributions are derived within SCET and are verified by the full QCD calculations up to $\mathcal{O}(\alpha_s^2)$. The corresponding collinear logarithms are resummed to all orders with the accuracy of NLL (${\cal{O}}(\alpha_s^n L^{n-1})$), while the transverse momentum-dependent logarithms are resummed to all orders with the accuracy of N$^3$LL for the unpolarized distribution and
N$^2$LL for the Sivers asymmetry.
\end{abstract}

%\date{\today}    
\maketitle

\section{Introduction}
Understanding the internal structure of hadronic events in Quantum Chromodynamics (QCD) continues to be a central objective in high-energy physics. Over the past few decades, significant progress has been made in mapping the three-dimensional momentum space of nucleons through transverse momentum-dependent distributions (TMDs). Many of the associated distributions such as the Sivers function, which describes the correlation between the transverse momentum of quarks and the transverse spin of the nucleon, play a central role in unraveling the dynamics of orbital angular momentum and spin physics. The upcoming Electron-Ion Collider (EIC)~\cite{AbdulKhalek:2021gbh} will provide an unprecedented precision laboratory to explore these multi-dimensional landscapes with high luminosity.

Over the past few years, the development of energy-correlator observables~\cite{Basham:1978zq,Basham:1978bw,Basham:1977iq,Ore:1979ry,Sveshnikov:1995vi,Korchemsky:1997sy,Korchemsky:1999kt,Belitsky:2001ij,Lee:2006nr,Hofman:2008ar,Moult:2025nhu} has offered a promising new perspective for probing nucleon structures~\cite{Gao:2025evv,Song:2025bdj,Kang:2026pro}. By tracking the angular distributions of energy deposited in the final state, these observables resolve complex QCD dynamics without traditional reliance on jet clustering. When applied to deep-inelastic scattering (DIS), their behavior spans two distinct physical limits. In the back-to-back (current fragmentation) kinematic regime, the resulting angular correlations are closely tied to the transverse-momentum-dependent (TMD) recoil effects, establishing a natural bridge to standard TMD factorization techniques~\cite{Moult:2018jzp,Kang:2023big,Kang:2024dja,Bhattacharya:2025bqa,Gao:2025cwy,Fu:2025hpc}. Conversely, within the forward (target fragmentation) region, they bypass individual hadronization stages to offer a direct probe of spin-dependent partonic motion inside the nucleon~\cite{Liu:2022wop,Cao:2023oef,Liu:2024kqt,Chen:2024bpj,Zhu:2025qkx,Gao:2025cwy}.

In this work we introduce a new variant of such observables which focuses on the one-point charge correlator (QC) in lepton-nucleon collisions. We define the charge correlator as the total charge deposited at a given polar angle with respect to the incoming proton direction:
\begin{equation} \label{eq:definition-DIS}
    \Sigma(\theta)= \sum_i \int\, {d\sigma_{e p \to e+ i+X}}\, Q_i\, \Theta(\theta-\theta_{i})\,.
\end{equation}
Here, $\theta_{i}$ denotes the polar angle of the final-state hadron $i$ measured with respect to the incoming proton in the Breit frame, and $Q_i$ is its electric charge. Driven by the high-precision tracking capabilities at the future EIC, this observable can be measured very precisely on charged tracks, showing great potential for phenomenological studies.

\begin{figure}[htbp]
\begin{center}
\includegraphics[width=0.5\textwidth]{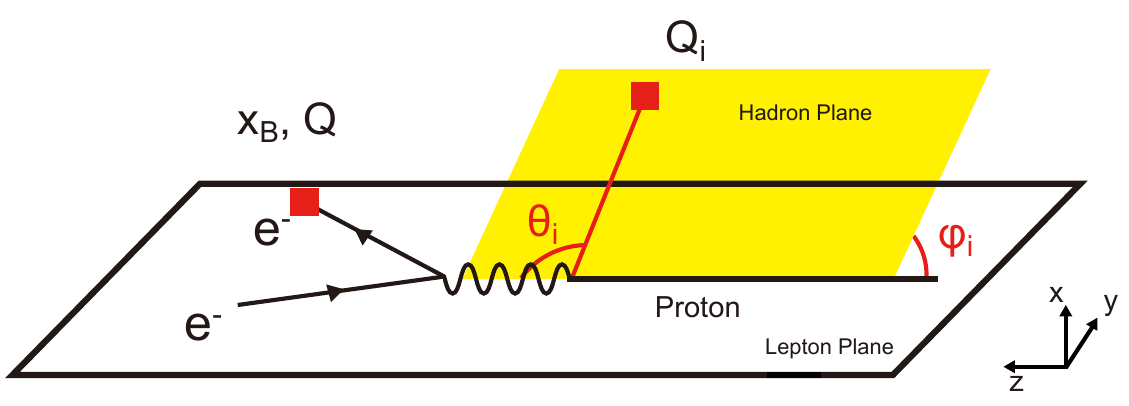}
\caption{Sketch of the one point charge correlator measurement showing the angle and the intuitive picture of the charge measurement operator.}
\label{fg:obs}
\end{center}
\end{figure}

To access the Sivers asymmetry, we further construct an azimuthal-angle-dependent version:
\begin{equation}
 \label{eq:definition-DISazi}
    \Sigma(\theta,\phi)= \sum_i \int\, {d\sigma_{e p \to e+ i+X}}\, Q_i\, \Theta(\theta-\theta_{i})\delta(\phi-\phi_i)\,,
\end{equation}
where $\phi_i$ is the azimuthal angle with respect to the lepton plane. A graphical illustration of the QC in DIS is provided in Fig.~\ref{fg:obs}.
The QC represents a generalization of the nucleon energy correlator (EC)~\cite{Liu:2023aqb} to the charge sector. For comparison, the EC is defined as:
\begin{equation}
 \label{eq:definition-EC}
    \Sigma_{EC}(\theta,\phi)= \sum_i \int\, {d\sigma_{e p \to e+ i+X}}\, \frac{E_i}{E_p}\, \Theta(\theta-\theta_{i})\delta(\phi-\phi_i)\,,
\end{equation}
where $E_i$ is the energy of the observed particle and $E_p$ is the proton energy.

The detector operation of charge correlator can be expressed formally in terms of QFT operators with intuitive space-time definitions \cite{Ore:1979ry,Sveshnikov:1995vi,Korchemsky:1997sy,Korchemsky:1999kt,Belitsky:2001ij, Lee:2006nr,Hofman:2008ar,Chicherin:2020azt,Cuomo:2025pjp}:

\begin{align} \label{Q-flow}
\mathcal Q(\vec n) = \int_0^\theta \sin \theta' d\theta'\lim_{r\to\infty} r^2\, \int_{-\infty}^\infty dt \,  n^i\, J_{i }(t,r\vec n)\,.
\end{align}
The theoretical structure of the charge correlator takes on different forms depending on the scaling of the angle $\theta$ defined above. In the target fragmentation region (TFR), where outgoing particles are emitted along the incoming hadron beam, this observable is theoretically described by a one-point charge correlator which provides a probe of the intrinsic nucleon dynamics. We demonstrate here the factorization into the partonic DIS cross section ${\hat \sigma}$ and a nonperturbative nucleon charge correlator (NQC):
\bea
\Sigma(\theta) = \int^1_{x_B}\frac{dz}{z} {\hat \sigma}_i\left(\frac{x_B}{z}\right) f_{i,{\rm QC}}(z,\theta) \,. 
\eea 
This factorization allows a direct connection between measured charge distributions and the underlying partonic structure of the nucleon.

In the current fragmentation region (CFR), where particles propagate in the opposite direction from the incoming nucleon, this observable can be used to extract the conventional transverse-momentum dependent parton distribution functions (TMDPDFs), the TMD fragmentation functions (TMDFFs) and the Sivers function. We perform the corresponding TMD resummation up to N$^3$LL accuracy for the unpolarized target and N$^2$LL accuracy for the study of the Sivers asymmetry.  

In both the TFR and TMD regimes, the observable can be systematically analyzed using a factorized framework based on soft-collinear effective theory (SCET)~\cite{Bauer:2000ew,Bauer:2000yr,Bauer:2001ct,Bauer:2001yt,Bauer:2002nz,Beneke:2002ph}. We validate our approach by comparing the singular distributions at $\mathcal{O}(\alpha_s)$ with those obtained from the factorized formula. The resummed distributions are then matched with fixed-order NLO QCD results for EIC kinematics. We briefly discuss non-perturbative effects in the CFR.

The remainder of the paper is organized as follows. In Sec. \ref{sec:kin} we introduce the kinematics in both the TFR and the CFR. In Sec.~\ref{sec:obs}, we demonstrate infrared safety of the observable. Sec.~\ref{sec:TFR} presents the factorization formulae in the TFR, while Sec.~\ref{sec:CFR} presents the factorization formulae in the TMD region.  The numerical consequence of the resummation, the fixed order $\alpha_s^2$ $\theta$-distribution and {\tt PYTHIA}~\cite{Sjostrand:2014zea, Bierlich:2022pfr} simulations are studied in Sec.~\ref{sec:num}. We conclude in Sec.~\ref{sec:con}.

\section{DIS kinematics in the Breit frame}
\label{sec:kin}

We begin by reviewing the kinematics of DIS and the momentum regions relevant to our study. The incoming and outgoing lepton momenta are denoted by $l^\mu$ and $l'^\mu$, while the initial proton momentum is denoted as $P^\mu$. The momentum carried by the virtual photon is $q^\mu \equiv l^\mu-l'^\mu$.
The standard Lorentz-invariant DIS variables are defined as
\begin{equation}
Q^2 \equiv -q^2, \qquad
x_B \equiv \frac{Q^2}{2P\cdot q},\qquad
y \equiv \frac{P\cdot q}{P\cdot l},
\end{equation}
where $Q^2$ characterizes the photon virtuality, $x_B$ is the Bjorken scaling variable, and $y$ is the inelasticity parameter.

Our discussion will be carried out in the Breit frame, in which the virtual photon carries momentum only along the $z$ axis:
\bea
q^{\mu}=\frac{Q}{2}(\bar{n}^\mu-n^\mu)=Q(0,0,0,-1)\,,
\eea
with the light-like vectors $\bar{n}^\mu=(1,0,0,-1)$ and $n^\mu=(1,0,0,1)$. The proton momentum is then given by
\bea
P^{\mu}=\frac{Q}{2x_B}n^\mu=\frac{Q}{2x_B}(1,0,0,1)\,.
\eea
We employ the usual light-cone notation $p^+\equiv \bar{n}\cdot p$ and $p^-\equiv n\cdot p$, such that a generic four-vector is written as $p^\mu=(p^+,p^-,\mathbf{p}_T)$. The polar angle of a final-state particle $i$ is defined by $\tan\theta_i=p_{i,T}/p_{i,3}$.

In the Breit frame, final-state radiation can be organized into two hemispheres. Particles emitted along the positive-$z$ direction define the target fragmentation region (TFR), while those emitted along the negative-$z$ direction correspond to the current fragmentation region (CFR). In the CFR, the detected hadrons originate primarily from the fragmentation of the struck parton. Depending on the angular configuration, the CFR can be further separated into two regimes. In the back-to-back (TMD) limit, $\bar{\theta}\equiv\pi-\theta\ll 1$, the relevant momentum scaling is
\begin{equation}
p_i \sim Q(\bar{\theta}^2,1,\bar{\theta}),
\end{equation}
and the cross section is described by TMD factorization. When $\theta\sim{\cal O}(1)$ the relevant momentum scaling is
\begin{equation}
p_i \sim Q(1,1,1),
\end{equation}
with large transverse momentum generated by hard QCD radiation. This region is described by fixed-order collinear factorization. The TFR corresponds to the small-angle region $\theta\ll 1$, where the observed hadrons scale as
\begin{equation}
p_i \sim Q(1,\theta^2,\theta).
\end{equation}
In this case the dominant contribution arises from spectator partons in the incoming nucleon, which continue along the beam direction and hadronize.

\begin{figure}[htbp]
\begin{center}
\includegraphics[width=0.5\textwidth]{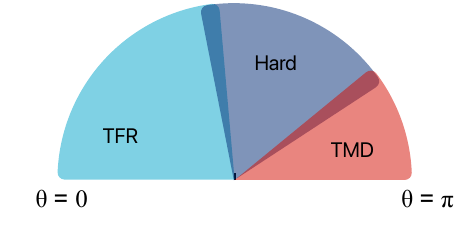}
\caption{Sketch of the relevant angular regions in the Breit frame.}
\label{fg:TFRCFR2}
\end{center}
\end{figure}

While the hard region is well described by fixed-order perturbative calculations, both the TFR and TMD regions involve large logarithmic contributions that must be resummed to obtain accurate predictions. This resummation, together with the corresponding factorization structure, can be systematically studied within the SCET framework. A schematic illustration of these angular regions in the Breit frame is shown in Fig.~\ref{fg:TFRCFR2}.

Two relevant planes can be defined in the Breit frame. The hadron plane is spanned by the final state hadron momentum and the proton beam direction, while the lepton plane is spanned by the incoming lepton momentum and proton beam directions. We denote the azimuthal angle between these two planes by $\phi_i$. In the following sections, we will study the azimuthal dependence of the charge correlator when the incoming proton is transversely polarized.

\section{Physics of charge correlators in the back-to-back limit}
\label{sec:obs}
\begin{figure}[htbp]
\begin{center}
\includegraphics[width=0.5\textwidth]{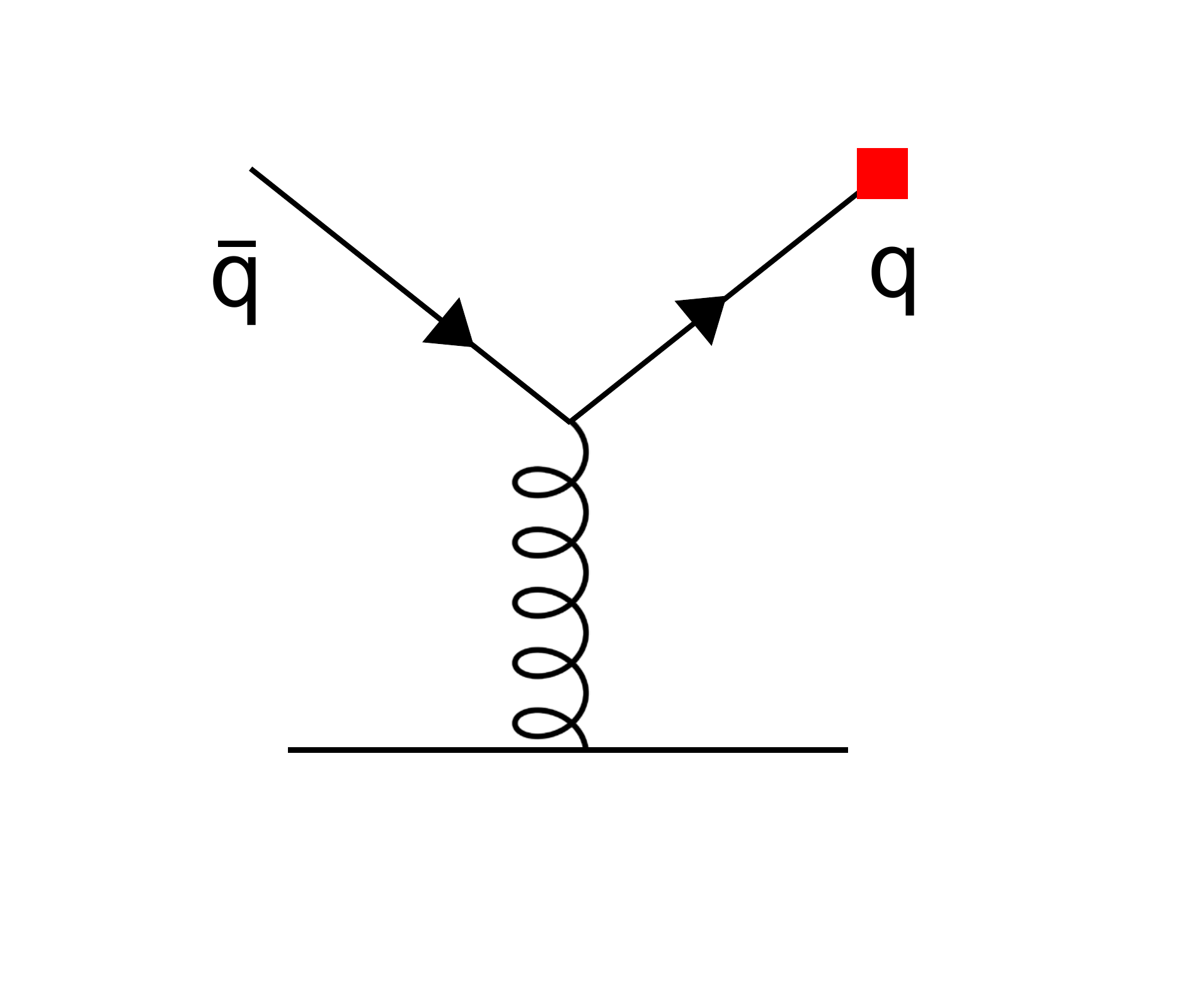}
\caption{Potential IRC-unsafe measurement of soft quark that originate from the branching of a
very soft gluon}
\label{fg:IR}
\end{center}
\end{figure}
The mechanism for infrared and collinear (IRC) safety of charge correlators differs from that of energy correlators, which we briefly discuss here. Beyond leading order in QCD perturbation theory, potential IRC unsafety stems from measuring quarks that originate from the splitting of a very soft gluon, as shown in Fig.~\ref{fg:IR}. 
 This process introduces soft charges into the final state in a manner that potentially disrupts the cancellations of real and virtual divergences necessary for IRC safety. In the case of energy correlators the vanishing of the individual energy of each emission in the soft limit leads to IRC safety. For QC, this cancellation occurs via charge conservation, after summing over 
 the $q$ and $\bar{q}$ states. For the potentially dangerous soft \(q\bar q\) contribution, the QCD soft kernel is symmetric under \(q\leftrightarrow\bar q\), whereas the signed charge insertion is antisymmetric. This ensures that perturbative methods can be safely applied to calculate the distribution.

\section{Factorization in the Target Fragmentation Region}
\label{sec:TFR}
Using the framework of SCET we now establish a factorization theorem for the QC in the TFR. The charge-weighted cumulant cross section $\Sigma(\theta)$ can be calculated as
\bea\label{eq:def-qc}
&& \Sigma(\theta) 
=  \frac{\alpha^2 }{Q^4}
\sum_{{\lambda=T,L} }
f_{\lambda} 
\epsilon^\ast_{\lambda,\mu} \epsilon_{\lambda,\nu}
\int 
d^4x
e^{i q \cdot x }
\langle P |j^{\mu\dagger}(x)
\hat{{\cal Q}}(\theta)
j^\nu  (0) | P \rangle 
\,, 
\eea 
where the dependence on $Q^2$ and $x_B$ is left implicit. $j^\nu$ denotes the electromagnetic current while $| P \rangle $ denotes the external proton state. The leptonic phase space measure can be expressed as
\bea
\frac{d^3 l}{(2\pi)^3\,2l^0}
=\frac{Q^2}{16\pi^2 s}\,dx_B\,dQ^2\,.
\eea
The polarization projectors of the virtual photon obey
\bea
\sum_{T=1,2}\epsilon^{\ast}_{T,\mu}\epsilon_{T,\nu}
&=&-g_{\mu\nu}+\frac{4x_B^2}{Q^2}P_\mu P_\nu\,,\nn\\
\epsilon^{\ast}_{L,\mu}\epsilon_{L,\nu}
&=&\frac{4x_B^2}{Q^2}P_\mu P_\nu\,,
\eea
up to contributions that vanish due to  gauge symmetry. $\epsilon^\mu_{T}$ and $\epsilon^\nu_L$ are the transverse and longitudinal polarization vectors of the virtual photon, respectively. The corresponding flux factors for these polarization states are
\bea\label{eq:flux}
f_T = 1-y+\frac{y^2}{2}\,,\qquad
f_L = 2-2y\,.
\eea
The operator $\hat{\cal Q}(\theta)$ measures the total charge accumulated inside a cone of opening angle $\theta$. It acts on a final state $|X\rangle$ as
\bea
\hat{{\cal Q}}(\theta) |X\rangle =
\sum_{i\in X} {Q_i} \Theta( \theta - \theta_i ) | X\rangle .
\eea
We note that if $\hat{\cal Q}(\theta)$ is replaced by the identity operator, Eq.~(\ref{eq:def-qc}) reduces to the standard inclusive DIS cross section.

We focus on the small-angle limit $\theta Q \ll Q$. The dominant contributions to $\Sigma(Q^2,\theta)$ arise from the following momentum regions:
\begin{itemize}

    \item hard modes with scaling $p_H=(p_H^+,p_H^-,p_{H,t})\sim Q(1,1,1)$;

    \item collinear modes scaling as $p_C\sim Q(1,\theta^2,\theta)$;

    \item soft modes with $p_S\sim Q(\theta^a,\theta^a,\theta^a)$ where $a\ge1$.
    
\end{itemize}
 In the small-angle limit, hard radiation is insensitive to the angular measurement since typical angles satisfy $\theta_H\sim{\cal O}(1)$, and thus do not contribute to the constraint imposed by $\Theta(\theta-\theta_{H})$. We therefore match the product $j^{\mu\dagger}\hat{{\cal Q}}(\theta)j^\nu$ onto a basis of SCET operators, consisting of quark and gluon contributions ${\cal O}_q$ and ${\cal O}_g$:
\bea
\langle P| j^{\mu\dagger}(x) \hat{{\cal Q}}(\theta) j^\nu(0) |P \rangle
= C_q^{\mu\nu} \langle P| {\cal O}_{q} |P \rangle + C_g^{\mu\nu} \langle P|{\cal O}_g |P \rangle ,,
\eea
where the $C_{q/g}^{\mu\nu}$ denote the corresponding hard matching coefficients. Note that the gluon coefficient $C_g^{\mu\nu}$ begins at ${\cal O}(\alpha_s)$. The SCET operators are defined as
\bea\label{eq:scetO} 
&&
{\cal O}_q(x,\theta) = 
{\bar \chi}_n(x)Y^\dagger(x) \frac{\gamma^+}{2}  \hat{{\cal Q}}(\theta) Y(0)\chi_n(0) \,,\nn \\  
&& 
{\cal O}_g(x,\theta) = 
{\cal B}_\perp (x){\cal Y}^\dagger(x)    \hat{{\cal Q}}(\theta) {\cal Y}(0){\cal B}_\perp (0) \,,
\eea 
written in terms of the standard collinear building blocks
\bea 
\chi_n(x) = W_n^\dagger(x) \xi_n(x) \,, \quad 
{\cal B}_{\perp}^\mu 
= \frac{1}{g_s}
[W_n^\dagger i {\cal D}_\perp^\mu W_n](x) \,.
\eea 
We note that both $\chi$ and ${\cal B}_\perp$ scale as $\theta$. Gauge invariance is ensured by the collinear Wilson line
\bea 
W_n(x) 
= \sum_{\rm perms} \exp 
\left(-\frac{g_s}{{\bar n}\cdot P_n} {\bar n}\cdot A_n(x)  \right)  \,, 
\eea 
which appears in the definitions of $\chi_n$ and ${\cal B}_\perp$. Soft interactions are encoded in the Wilson lines $Y$ and ${\cal Y}$ in the fundamental and adjoint representations, respectively. 
Since these Wilson lines are electrically neutral, the charge measurement commutes with them:
 \bea\label{eq:eycommute} 
[\hat{{\cal Q}},Y] = [\hat{{\cal Q}},{\cal Y}] = 0\,.
\eea 

With these ingredients, the hadronic matrix element in Eq.~(\ref{eq:def-qc}) is expressed in SCET as 
\bea 
&& \int 
d^4x
e^{i q \cdot x }
\,  
\langle P |j^{\dagger\mu}(x)    \, 
\hat{{\cal Q}}(\theta) \, 
j^\nu  (0) | P \rangle  \nn \\ 
&=& \int 
d^4x
e^{i q \cdot x }
\Bigg( C_q^{\mu\nu}(x) 
\langle P |
{\bar \chi}_n(x)Y^\dagger(x) \frac{\gamma^+}{2} \hat{{\cal Q}}(\theta) Y(0)\chi_n(0)
| P \rangle   \nn \\ 
&& 
+
 C_g^{\mu\nu}(x) 
\langle P| {\cal B}_\perp (x){\cal Y}^\dagger(x)  \hat{{\cal Q}}(\theta) {\cal Y}(0){\cal B}_\perp (0)  |P \rangle 
\Bigg)  \,,
\eea 
The momentum transfer satisfies $q\sim Q(1,1,1)$, implying that the Fourier integral is dominated by coordinate separations of order $x\sim Q^{-1}(1,1,1)$. Increasing $Q$ corresponds to probing the nucleon over shorter distances, as expected from the usual DIS interpretation. 

At leading power we expand the SCET operators in the light-cone coordinate $x^-$. We perform a multiple expansion of the collinear fields and the soft Wilson lines, obtaining
\bea 
&& \int 
d^4x
e^{i q \cdot x }
\,  
\langle P |j^{\dagger\mu}(x)    \, 
\hat{{\cal Q}}(\theta)\, 
j^\nu  (0) | P \rangle  \nn \\ 
&=& \int 
d^4x 
e^{i q \cdot x } 
\Bigg( C_q^{\mu\nu}(x) 
\langle P |
{\bar \chi}_n(x^-)Y^\dagger(0) \frac{\gamma^+}{2} \hat{{\cal Q}}(\theta) Y(0)\chi_n(0)
| P \rangle   \nn \\ 
&& +
 C_g^{\mu\nu}(x) 
\langle P| {\cal B}_\perp (x^-){\cal Y}^\dagger(0) \hat{{\cal Q}}(\theta) {\cal Y}(0){\cal B}_\perp (0)  |P \rangle 
\Bigg)  \,.
\eea 
Using the commutators in Eq.~(\ref{eq:eycommute}) together with $Y^\dagger Y={\cal Y}^\dagger{\cal Y}=1$, all soft Wilson lines cancel, and we obtain
\bea 
&& \int 
d^4x
e^{i q \cdot x }
\,  
\langle P |j^{\dagger\mu}(x)    \, 
\hat{{\cal Q}}(\theta) \, 
j^\nu  (0) | P \rangle  \nn \\ 
&=& \int 
d^4x
e^{i q \cdot x }  \Bigg( C_q^{\mu\nu}(x) 
\langle P |
{\bar \chi}_n(x^-) \frac{\gamma^+}{2} \hat{{\cal Q}}(\theta)   \chi_n(0)
| P \rangle   \nn \\
&& \hspace{5.ex} +
 C_g^{\mu\nu}(x) 
\langle P| {\cal B}_\perp (x^-)   
\hat{{\cal Q}}(\theta)  {\cal B}_\perp (0)  |P \rangle 
\Bigg)  \,.
\eea 
This makes manifest that in the small-$\theta$ limit, soft radiation is treated inclusively and does not generate logarithmic enhancements. Inserting this result back into Eq.~(\ref{eq:def-qc}) gives
\bea 
&& \Sigma(\theta)
=  \frac{\alpha^2 }{Q^4}
\sum_{{\lambda=T,L} }
f_{\lambda} 
\epsilon^\ast_{\lambda,\mu} \epsilon_{\lambda,\nu} \nn \\ 
&&   \times
\int 
d^4x
e^{i q \cdot x } \Bigg( C_q^{\mu\nu}(x) 
\langle P |
{\bar \chi}_n(x^-) \frac{\gamma^+}{2} \hat{{\cal Q}}(\theta)   \chi_n(0)
| P \rangle   \nn \\ 
&& 
\hspace{5.ex} +
 C_g^{\mu\nu}(x) 
\langle P| {\cal B}_\perp (x^-)  \hat{{\cal Q}}(\theta)  {\cal B}_\perp (0)  |P \rangle 
\Bigg)
\,.
\eea 
We insert a complete set of collinear final states,
\bea
1=\sum_{X_C}|X_C\rangle\langle X_C|\,,
\eea
into the hadronic matrix elements appearing in $\Sigma(\theta)$. 
We then translate along the $x^-$ direction. In particular, 
\bea
\langle P|\,\bar{\chi}_n(x^-)\,|X_C\rangle
&=&
\langle P|\,T^\dagger T\,\bar{\chi}_n(x^-)\,T^\dagger T\,|X_C\rangle \nn\\
&=&
\langle P|\,e^{iP^+\frac{x^-}{2}}\,
\bar{\chi}_n(0)\,
e^{-iP_C^+\frac{x^-}{2}}\,|X_C\rangle\,,
\eea
where $T$ denotes the translation operator in $x^-$, and $P_C^+$ is the large light-cone momentum component carried by the collinear radiation. 
Introducing the auxiliary identity
\bea
1=P^+\int dz\,\delta\!\left((1-z)P^+-P_C^+\right)\,,
\eea
to define the momentum fraction variable $z$, we obtain

\bea 
&& \Sigma(\theta)
=  \frac{\alpha^2 }{Q^4}
\sum_{{\lambda=T,L} } 
f_{\lambda} 
\epsilon^\ast_{\lambda,\mu} \epsilon_{\lambda,\nu} \nn \\ 
&&   \times
P^+ \int  d z \delta((1-z) P^+ - P_C^+)  \int d^4x 
e^{i q \cdot x } 
e^{i(P^+ - P_C^+ ) \frac{x^-}{2}} 
\nn \\ 
&&\times 
\Bigg( C_q^{\mu\nu}(x) 
\langle P |
{\bar \chi}_n(0) \frac{\gamma^+}{2} \hat{{\cal Q}}(\theta)   |X_C\rangle \langle X_C| \chi_n(0)
| P \rangle   \nn \\
&& \hspace{5.ex} +
 C_g^{\mu\nu}(x) 
\langle P| {\cal B}_\perp (0)   \hat{{\cal Q}}(\theta)  
|X_C\rangle \langle X_C|
{\cal B}_\perp (0)  |P \rangle 
\Bigg)
\,, 
\eea 
Using $P^\mu\simeq P^+ n^\mu/2$ up to ${\cal O}(\Lambda_{\rm QCD}/Q)$ corrections, $(P^+-P_C^+)\frac{x^-}{2}$ can be rewritten as $zP^+ \frac{x^-}{2} = z P\cdot x$. The delta function can be represented in Fourier space as 
$$
\delta((1-z)P^+ - P_C^+) = \int \frac{dy^-}{4\pi} e^{i[(1-z)P^+-P_C^+]\frac{y^-}{2}}.
$$
After shifting the collinear fields by $y^-$ and performing the $x$ integral, the cumulant cross section can be written into the factorized form
\bea\label{eq:eec-fact1}
  \Sigma(\theta)
&=&  
 \int  d z \Bigg(   
H_q(z,x_B,Q^2)   \, f_{q,{\rm QC}}(z,P^+\theta)  \nn \\
&&  +   H_g(z,x_B,Q^2)  \,
f_{g,{\rm QC}}(z,P^+\theta)
\Bigg)
\,,
\eea 
where the hard functions are given by
\bea 
&&
\hspace{-2.ex}
H_q = \frac{\alpha^2 }{Q^4}
\sum_{{\lambda=T,L} }
f_{\lambda} 
\epsilon^\ast_{\lambda,\mu} \epsilon_{\lambda,\nu}
 \int 
d^4x
e^{i (q+zP  ) \cdot x } 
 C_q^{\mu\nu}(x) P^+    \,,  \nn \\ 
&& 
\hspace{-2.ex}
H_g = \frac{\alpha^2 }{Q^4}
\sum_{{\lambda=T,L} }
f_{\lambda} 
\epsilon^\ast_{\lambda,\mu} \epsilon_{\lambda,\nu}
 \int 
d^4x
e^{i (q+zP  ) \cdot x } 
 C_g^{\mu\nu}(x)  \,. 
\eea 
The nucleon charge correlators are defined as
\bea\label{eq:fqx} 
&& f_{q,{\rm QC}}(z,P^+\theta) 
=  \int \frac{dy^-}{4\pi} e^{- i z P^+ \frac{y^-}{2} }  \langle P |
{\bar \chi}_n\left(\frac{y^-}{2}n^\mu\right) \frac{\gamma^+}{2} 
\hat{{\cal Q}}(\theta)  \chi_n(0)
| P \rangle  \,, 
\eea 
for quarks, and 
\bea\label{eq:fgx}
&& f_{g,{\rm QC}}(z,P^+\theta)  
= 
\int \frac{dy^-}{4\pi } e^{- i z P^+ \frac{y^-}{2} }  
P^+ 
\langle P| {\cal B}_\perp
\left(\frac{y^-}{2}n^\mu \right)   
\hat{{\cal Q}}(\theta)   
{\cal B}_\perp (0)  |P \rangle 
\eea 
for the gluon. 
These matrix elements provide the operator definitions of the quark and gluon  nucleon charge correlators in momentum space. 

The hard functions $H_q$ and $H_g$ can be fixed as follows.
\begin{itemize}
\item Replacing the measurement operator $\hat{\mathcal{Q}}(\theta)$ by the identity $1=\sum_X |X\rangle\langle X|$ in the QC matrix elements restores the standard operator definition of the collinear PDFs. In this limit, the corresponding observable reduces to the usual inclusive DIS cross section.

\item The hard matching coefficients are insensitive to the structure of the collinear sector. As a result, they are unaffected by whether the collinear matrix elements contain the operator $\hat{\mathcal{Q}}(\theta)$ or the identity operator.

\end{itemize}
Consequently, we identify
\bea 
H_q = \frac{1}{z} \, {\hat \sigma}_{q}\left(\frac{x_B}{z},Q^2\right)\,, \quad 
H_g = \frac{1}{z} \, {\hat \sigma}_{g}\left(\frac{x_B}{z},Q^2\right)\,, 
\eea 
where $\hat{\sigma}_{q,g}$ are the standard partonic DIS cross sections. The factorization theorem takes the compact form
\bea\label{eq:fact-x}
\Sigma(\theta)
&=&  \sum_{i= q,g}  \int \frac{dz}{z} \hat{\sigma}_i\left(\frac{x_B}{z},Q^2 \right) f_{i,{\rm QC}}(z,P^+\theta) \,. 
\eea 
This form makes it explicit that all $\theta$ dependence resides in the collinear correlators $f_{i,{\rm QC}}$, so that the angular cumulant directly probes the nucleon charge correlators.

\subsection{Matching onto the Collinear PDF when $\theta Q \gg \Lambda_{\rm QCD}$}\label{sec:pertmatch}

When $\theta Q \gg \Lambda_{\rm QCD}$, 
the collinear modes admit a further separation into a hard-collinear mode $C_1$ with scaling $p_{C_1}\sim Q(1,\theta^2,\theta)$ and a $C_2$ mode in SCET$_{\rm II}$ with $p_{C_2}\sim Q(1,\lambda^2,\lambda)$, where $\lambda \equiv \Lambda_{\rm QCD}/Q \ll \theta$. The SCET operators in Eq.~(\ref{eq:scetO}) can be further matched onto the SCET$_{\rm II}$ operators such that 
\bea\label{eq:ItoII} 
{\cal O}_{i}(x^-) = \sum_{j=q,g}
C_{j}(x^-)
{\cal O}_{j,{\rm II}}(x^-) \,.
\eea 
The ${\cal O}_{j,{\rm II}}$ have the same Dirac and color structure as ${\cal O}_j$ but are constructed solely from $C_2$ fields and do not contain the measurement operator $\hat{\cal Q}(\theta)$.

It is helpful to examine the action of $\hat{\cal Q}(\theta)$ on a generic state $|X\rangle=|X_{C_1},X_{C_2}\rangle$. From the definition, we have 
\bea
&&\hat{\cal Q}(\theta)|X_{C_1},X_{C_2} \rangle =
\sum_{
\substack{i\in X_{C_1}\\
j\in X_{C_2} }}
\left( Q_i\Theta(\theta-\theta_i ) 
+ Q_j  \Theta(\theta-\theta_j ) 
\right)
|X \rangle  \,.  
\eea 
Since emissions in the $C_2$ sector satisfy $\theta_j\sim\lambda\ll\theta$, the corresponding step functions reduce to unity. The above expression can then be rewritten as
\bea\label{eq:c1c2} 
&&\hat{\cal Q}(\theta)|X_{C_1},X_{C_2} \rangle 
=\left(
\sum_{i\in X_{C_1}}
 - Q_i\Theta(\theta_i-\theta ) 
+ Q_X   
\right)
|X_{C_1},X_{C_2} \rangle \,\, \,   \,.  
\eea 
with 
\begin{equation}
Q_X\equiv \sum_{
\substack{i\in X_{C_1}\\
j\in X_{C_2} }}(Q_i+Q_j)
\end{equation}
and where $\Theta(\theta-\theta_i)=1-\Theta(\theta_i-\theta)$ has been used. The $Q_X$ term in Eq.~(\ref{eq:c1c2}) acts on both $C_1$ and $C_2$ modes simultaneously and contributes to the $f_{\rm QC}$ so that
\bea 
&& f_{i,{\rm QC}} 
\supset (Q_P - Q_i)f_i(z) \,, \qquad \text{with $i=q,g$\,,}
\eea 
where $f_i(z)$ is the collinear PDF. The $- Q_i\Theta(\theta_i-\theta)$ term in Eq.~(\ref{eq:c1c2}) acts only on the $C_1$ modes. In the matching onto SCET$_{\rm II}$, it combines with the Wilson coefficient $C_j(x^-)$ in Eq.~(\ref{eq:ItoII}) and generates the corresponding perturbative matching kernel. Its contribution can be written as
\bea 
f_{i,{\rm QC}} \supset - \sum_j \int^1_z \frac{d\xi}{\xi} I'_{ij}\left(\frac{z}{\xi} ,\theta\right)\, \left[ f_j(\xi) \right] \,. 
\eea 
Here $I_{ij}'$ is the matching coefficient. It can be calculated perturbatively and starts at ${\cal O}(\alpha_s)$.  

Collecting all contributions, the matched expression in the regime $\theta Q \gg \Lambda_{\rm QCD}$ becomes
\bea\label{eq:Ix} 
&&f_{i,{\rm QC}}(z,\theta)= Q_Pf_i(z) - \int_z^1 \frac{d \xi}{\xi} 
I_{ij}\left(
\frac{z}{\xi}, \theta 
\right) \,   f_j(\xi) \,, 
\eea 
where $I_{ij}(z) = Q_i\delta(1-z) + I'_{ij}(z)$. 
\subsection{NLO Matching Coefficient for $I_{ij}$}\label{sec:collinear}

The matching coefficients $I_{ij}$ in Eq.~(\ref{eq:Ix}) can be calculated by comparing the partonic matrix elements of the QC observable in Eqs.~(\ref{eq:fqx}) and (\ref{eq:fgx}) with the collinear PDF matrix elements within SCET. We can replace the external hadronic states $|P\rangle$ and $|X\rangle$ with on-shell quark and gluon states, allowing a perturbative calculation. Working in dimensional regularization, higher order corrections to the collinear PDF are scaleless and vanish identically. The matching is fully determined by the partonic matrix elements of the QC operator in Eqs.~(\ref{eq:fqx}) and (\ref{eq:fgx}).
 
At NLO, the contribution of the second term in Eq.~(\ref{eq:Ix}) is given by
\bea 
&&f_{i,{\rm QC}} 
= - P^+  \int d\xi  \, 
\delta((\xi - z)P^+ - g^+) 
\nn \\ 
&\times & \int \frac{d^d p}{(2\pi)^{d-1}}
\delta(p^2) \, 
\left(Q_j-Q_i \right)  \,
\Theta(\theta_p - \theta) \, 
\nn \\ 
&\times &
(8\pi\alpha_s)  \mu^{2\epsilon}
\frac{1-\frac{z}{\xi}}{p_t^2} P^{(0)}_{ij}
\left( \frac{z}{\xi},\epsilon \right)
f_j(\xi)   \,, 
\eea 
where $p^\mu$ is the momentum of the detected parton, and $p_t$ is its transverse component. $\xi$ is the momentum fraction carried by the incoming parton. 
Here $P^{(0)}_{ij}$ are the ${\cal O}(\alpha_s)$ splitting kernels. We parameterize the phase space as 
\bea 
 \frac{d^dp}{(2\pi)^{d-1}}\delta(p^2) 
= \frac{1}{16 \pi^2} 
\frac{(4\pi)^\epsilon}{\Gamma(1-\epsilon)}
\frac{dp^+}{p^+}
\left( \frac{p^+}{2}\right)^{2-2\epsilon}
d\theta_p^2  
\theta_p^{-2\epsilon} 
\,, 
\eea 
where we have used $p_t = \theta_p \frac{p^+}{2}$. 
At NLO, the only non-vanishing contribution is 
\bea 
f_{i,{\rm QC}} (z,\theta)
&= &(Q_i-Q_j) \frac{\alpha_s }{2 \pi \epsilon} \left(\frac{z Q\theta}{2x_B}\right)^{-2\epsilon}
\frac{(4\pi\mu^{2})^\epsilon}{\Gamma(1-\epsilon)} \int_{z}^1 d\xi  
\frac{( \xi/z-1)^{-2\epsilon}}{\xi}
 P^{(0)}_{ij}
\left( \frac{z}{\xi},\epsilon \right)
f_j(\xi) 
\eea
where~\cite{Catani:1998nv}
\bea\label{eq:split}
&& P^{(0)}_{qq}(z,\epsilon) = C_F \left( \frac{1+z^2}{1-z}
-\epsilon(1-z) \right) \,, \nn \\ 
&& P^{(0)}_{gq}(z,\epsilon) = 
C_F
\left( \frac{1+(1-z)^2}{z} -\epsilon z
\right) \,, \nn \\ 
&& P^{(0)}_{qg}(z,\epsilon) = 
T_R
\left( z^2+(1-z)^2 - 2\epsilon z(1-z)  
\right) \,, \nn \\ 
&& P^{(0)}_{gg}(z,\epsilon) =  
2C_A\left(
\frac{z}{1-z}
+ \frac{1-z}{z}
+z(1-z) 
\right)\,.
\eea
The unrenormalized NLO matching coefficient is
\bea 
I_{ij}^{un,(1)}\left(\frac{z}{\xi},\theta\right) &=& (Q_j-Q_i)\frac{\alpha_s }{2 \pi \epsilon} \left(\frac{zQ\theta}{2x_B}\right)^{-2\epsilon}
\frac{(4\pi\mu^{2})^\epsilon}{\Gamma(1-\epsilon)}   
( \xi/z-1)^{-2\epsilon}
 P^{(0)}_{ij}
\left( \frac{z}{\xi},\epsilon \right)
\nn\\
&=&(Q_j-Q_i) \frac{\alpha_s }{2 \pi } \left(\ln\frac{4x_B^2\mu^{2}}{z^2Q^2\theta^2}+\frac{1}{\epsilon}\right)
\frac{(4\pi)^\epsilon}{\Gamma(1-\epsilon)}   
 P^{(0)}_{gq}
\left( \frac{z}{\xi} \right)+(Q_j-Q_i) \frac{\alpha_s }{2 \pi }d_{gq}\left(\frac{z}{\xi} \right) \nonumber \\
\eea 
where
\bea 
d_{ij}\left(z \right)=2\ln\left( \frac{z}{1-z}\right) p^{(0),0}_{ij} \left(z \right)+p^{(0),1}_{ij}\left(z \right).
\eea 
Here, the
 $p_{ij}^{(0),k}(z)$ are the coefficients of  $\epsilon^k$ with $k=0,1$ in the splitting kernels $P_{ij}^{(0)}(z,\epsilon)$ of Eq.~(\ref{eq:split}).  
The NLO  renormalized matching coefficient is then 
\bea\label{eq:Iij}
I_{ij}^{(1)} \left(\frac{z}{\xi},\theta\right)
&=& (Q_j-Q_i) \frac{\alpha_s }{2 \pi } \left[\ln\frac{4x_B^2\mu^{2}}{z^2Q^2\theta^2}
 P^{(0)}_{ij}
\left( \frac{z}{\xi} \right)+d_{ij}\left(\frac{z}{\xi} \right)\right]
\nn
\\
&=& \frac{\alpha_s }{2 \pi } \left[\ln\frac{4x_B^2\mu^{2}}{z^2Q^2\theta^2}
 P^{Q,(0)}_{ij}
\left( \frac{z}{\xi} \right)+d^{Q}_{ij}\left(\frac{z}{\xi} \right)\right] \, ,
\eea 
where $P_{ij}^{(0)}(z)$ are the ${\cal O}(\alpha_s)$ splitting functions
\bea 
&&P_{qq}^{(0)}(z) = C_F \left(\frac{1+z^2}{1-z} \right)_+\,, \nn \\ 
&&P_{gq}^{(0)} = P_{gq}^{(0)}(z) = C_F \frac{1+(1-z)^2}{z} \,, \nn \\ 
&&P_{qg}^{(0)}(z) = T_R (z^2+(1-z)^2) \,, \nn \\ 
 &&P_{gg}^{(0)}(z) = 2C_A\left( 
\frac{z}{(1-z)_+} + \frac{1-z}{z}
+z(1-z) 
\right) + \frac{\beta_0}{2}\delta(1-z) \,. 
\eea 
and we defined
\bea\label{eq:pq} 
P^Q_{ij}\equiv (Q_j-Q_i)P_{ij} \,,
\nn
\\
d^Q_{ij}\equiv (Q_j-Q_i)d_{ij}\,,
\eea
We note that unlike the energy correlator in~\cite{Cao:2023oef}, $I_{qq}$ and $I_{gg}$ vanish at this order because of the factor $(Q_j-Q_i)$. 

\subsection{Evolution Equations} 
\label{sec:evolution}

When $\alpha_s \ln \theta^2 \sim 1$, logarithmic enhancements become large and must be resummed to all orders. The NLO results obtained in Sec.~\ref{sec:collinear} provide the ingredients necessary to achieve next-to-leading logarithmic (NLL) accuracy, resumming terms of the form $\alpha_s^k \ln^{k}\theta^2$ and $\alpha_s^k \ln^{k-1}\theta^2$.

From the consistency relation  
\bea 
\frac{d}{d\ln\mu^2} \Sigma(\theta) = 0  \,, 
\eea 
we deduce that the QC satisfies the  DGLAP evolution equation
\bea\label{eq:evo1} 
&& \frac{d}{d\ln\mu^2} f_{i,{\rm QC}}(z,\theta) = \sum_j \int_z^1 \frac{d\xi}{\xi}  P_{ij}\left(\frac{z}{\xi}\right)
 f_{j,{\rm QC}}(\xi,
\theta)\,.
\eea 
When $\theta P^+ \gg \Lambda_{\rm QCD}$, as shown above, the $f_{QC}$ can be matched onto the collinear PDFs, with all $\theta$ dependence occurring only in the perturbative matching coefficients $I$. Since $f_{\rm QC}$ is dimensionless and  the matching coefficient is insensitive to the hadronic state, the $P^+\theta$ will appear only in the form  $\ln\frac{z P^+\theta}{\mu}$. $\Sigma$ can therefore also be written as
\bea
\label{eq:fact-N}
\Sigma(\theta)= \sum_{i= q,g} \int \frac{dz}{z} \hat{\sigma}_i\left(\frac{x_B}{z} \right) f_{i,{\rm QC}}\left(z,\ln \frac{zQ\theta}{x_B\mu}\right) \,. 
\eea 
We have used that $P^+ = \frac{Q}{x_B}$ in the Breit frame. We note that $\mu$-dependence also enters through the strong coupling and the collinear PDFs. This expression is consistent with the NLO result presented above.

The renormalization group equation governing the coefficient function follows directly from Eqs.~(\ref{eq:evo1}) and (\ref{eq:Ix}):
\begin{align}
\label{eq:match-rg}
&\frac{d {I}_{ij}}{d \ln \mu^2}
\left(
z\,,
\ln \frac{z Q\theta}{x_B\mu}
\right)
= 
\int_z^1 
\frac{d \xi}{\xi}
\bigg[
 P_{ik}(\xi)
{I}_{kj}\left(
\frac{z}{\xi}\,,
\ln \frac{z Q\theta}{\xi x_B\mu}
\right)
-
{I}_{ik}\left(
\frac{z}{\xi}\,,
\ln \frac{z Q\theta}{\xi x_B\mu}
\right)
 P_{kj}(\xi)
\bigg]
\,.
\end{align}
The first term in Eq.~(\ref{eq:match-rg}) originates from the evolution of the QC,
while the second term arises from the evolution of PDF.
The solution  up to NNLO reads
\begin{align}
\label{eq:RG-solution}
{I}_{ij}&
\left(
z\,,
\ln \frac{Q\theta}{x_B\mu}
\right)
=
Q_i\delta_{ij}\delta(1-z)
+
a_s(\mu)
\bigg( 
{I}^{(1)}_{ij}(z)
\nn\\
-&
L_q
P_{ij}^{Q,(0)}(z)
\bigg)
+
a_s^2(\mu)
\bigg(
{I}^{(2)}_{ij}(z)
+
L^2_q
\nn\\
\times
&
\bigg[
\frac{P_{ik}^{(0)}\otimes P_{kj}^{Q,(0)}(z)}{2}
-
\frac{P_{ik}^{Q,(0)}\otimes P_{kj}^{(0)}(z)}{2}
\nn\\
+&
\frac{1}{4}\beta_{0}
P_{ij}^{Q,(0)}(z)
\bigg]
+L_q
\bigg[
{I}^{(1)}_{ik}
\otimes
P_{kj}^{(0)}(z)
\nn\\
-
&
P_{ik}^{(0)}
\otimes
{I}^{(1)}_{kj}(z)
-
\frac{\beta_0}{2} 
{I}^{(1)}_{ij}(z)
-
P_{ij}^{Q,(1)}(z)
\nn\\
+&
2
\int \frac{d\xi}{\xi} \ln \xi P_{ik}^{(0)}(\xi)P_{kj}^{Q,(0)}\left(\frac{z}{\xi}\right)
% \left[\ln z \,P_{ik}^{(0)}(z)\right]
% \otimes
% P_{kj}^{Q,(0)}(z)
\bigg]
\bigg)
+\mathcal{O}(a_s^3(\mu))
\,,
\end{align}
where $a_s(\mu)=\alpha_s(\mu)/(2\pi)$ and $L_q=2\ln \frac{zQ\theta}{2x_B\mu}$.

A convenient implementation of the resummation is to compute the partonic cross section $\hat{\sigma}_{c,\lambda}$ together with the collinear PDFs $f_i$ at a hard scale $\mu \sim Q$, while evolving the matching coefficient $I$ from the natural scale $\mu_0 \sim Q\theta$ up to $\mu \sim Q$ using the evolution equation in Eq.~(\ref{eq:match-rg}). An equivalent formulation is to define both the PDFs $f_i$ and the matching coefficients $I_{ij}$ at the scale $\mu_0 \sim Q\theta$, and evolve the full QC distribution to the hard scale using the evolution equation in Eq.~(\ref{eq:evo1}). This evolution follows the standard DGLAP structure and can be implemented numerically using {\tt HOPPET}~\cite{Salam:2008qg} or {\tt APFEL}~\cite{Bertone:2013vaa,Bertone:2017gds}.

\subsection{Resummation in Mellin space}\label{sec:nll}

We note that an iterative Mellin-space solution strategy developed for nucleon energy correlators~\cite{Cao:2023oef} can be applied to derive an analytic solution for the QC as well. We use this approach as a check on our numerical implementation. The Mellin-transformed observable is defined as
\bea\label{eq:qc-def-melliin} 
\Sigma_N(\theta) = \sum_i \int d\sigma(x_B,Q^2,p_i) x_B^{N-1} Q_i
\Theta(\theta - \theta_i) \,. 
\eea 
Using the factorized form in Eq.~(\ref{eq:fact-x}), we obtain
\begin{align}
\label{eq:fac-N-space}
\Sigma_N(\theta) 
\simeq&
 \frac{2\pi\alpha_e^2}{Q^4}
       \int_0^1 d u \, u^{N-1-n}
  \sum_{\Gamma=T,L}
    \sum_{n=0}^2
    f^n_\lambda
  \bigg[
  \sum_i Q_i^2\hat{\sigma}_{i}^\lambda(u) 
   f_{i,\rm QC}\left(N-n,\ln\frac{Q\theta}{2 u \mu}\right)   
   \nn\\
  +  & \frac{1}{N_f} \sum_{i=1}^{N_f} Q_i^2 
  \hat{\sigma}_{g}^\lambda(u)    f_{g,\rm QC}\left(N-n,\ln\frac{Q\theta}{2u \mu}\right)
  \bigg]
  \,,
\end{align}
where $u = \frac{x_B}{z}$. The approximate equality arises from extending the lower kinematic limit of $x_B$ down to zero. Such effects are suppressed in Mellin space due to the presence of the weight $x_B^N$. The Mellin space function $f_{i,{\rm QC}}(N,\ln\frac{Q\theta}{u\mu})$ is defined as
\bea\label{eq:mellin-s} 
f_{i,{\rm QC}}(N,\ln\frac{Q\theta}{u \mu})
= \int_0^1 dz\, z^{N-1} f_{i,{\rm QC}}(z,\ln\frac{ Q\theta}{ z u \mu})\,. \,\, 
\eea 
The additional decomposition of the flux,
$f_\lambda=\sum_{n=0}^2 f_\lambda^n x_B^{-n}$,
is required due to the relation $y = Q^2/(x_B s)$. In Mellin space, Eq.~(\ref{eq:evo1}) becomes
\bea \label{eq:evo2}
&& \frac{d}{d\ln\mu^2} f_{i,{\rm QC}}(N,\ln\frac{Q\theta}{u \mu}) \nn \\
&=&
\sum_j \int d\xi \xi^{N-1} 
P_{ij}\left(\xi\right)
 f_{j,{\rm QC}}(N,\ln\frac{Q\theta}{\xi \, u \mu}) 
 \,.
\eea
Following the iterative solution strategy developed for nucleon energy correlators~\cite{Cao:2023oef}, and detailed in Appendix~\ref{sec:derive}, the NLL-resummed solution can be written in a compact form as 
\bea\label{eq:fNLLevo} 
&& f_{i,{\rm QC}}(N,\ln\frac{Q\theta}{u\mu}) = f_i(N,\mu)  -    
   {\cal D}_{ik}^N(\mu,\mu_0)
  \, I_{kj}(N,\ln\frac{Q\theta}{u\mu_0}) f_{j}(N,\mu_0) \nn \\ 
 && - 
 \frac{\alpha_s(\mu_0)}{2\pi}
 {\cal N}_{ik} 2P_{kj}^{Q,(0)}(N)f_j(N,\mu_0) \,. \nn \\ 
\eea 
Here $I_{ij}(N,\ln\frac{Q\theta}{u\mu_0})$ is the NLO matching coefficient in Eq.~(\ref{eq:Iij}) evaluated at the scale $\mu_0$ in Mellin space, and the evolution factor ${\cal D}^N_{ij}(\mu,\mu_0)$ is  the DGLAP evolution in  Mellin space, 
\bea\label{eq:D}
{\cal D}^N_{ij}(\mu,\mu_0) = 
\exp\left[\int_{\mu_0}^{\mu} d\ln\mu^2 P(N,\mu)\right]_{ij} \,.
\eea 
To realize  NLL resummation, we need $P_{ij}(N)$ through NLO in the evolution factor ${\cal D}^N_{ij}$. The correction to the DGLAP evolution starts from $\alpha_s^nL^{n-1}$ order, in which  
\bea
{\cal N}_{ij}
&=&  \int_{\mu_0}^\mu d\ln{\mu_1^2}
{\cal D}_{ik}^N(\mu,\mu_1) \tilde{P}_{kl}(N,\mu_1)
{\cal D}_{lj}^N(\mu_1,\mu_0) \,. \quad 
\eea 
We note that at NLL both ${\cal D}$ and ${\cal N}$ can be integrated analytically.
 Including the evolution of the PDFs  to the hard scale,
$f_j(N,\mu_0) = {\cal D}^{-1}_{jk}(N) f_k(N,\mu)$,
we obtain the corresponding evolution of the matching coefficient,
\bea\label{eq:INLLevo} 
 I_{ij}(N,\ln\frac{Q\theta}{u\mu}) &=&   
  {\cal D}^N_{ik}(\mu,\mu_0)
  \, I_{kl}(N,\ln\frac{Q\theta}{u\mu_0}) {{\cal D}_{lj}^N}(\mu_0,\mu) \nn \\ 
 &&  
 \hspace{-10.ex} 
+  \frac{\alpha_s(\mu_0)}{2\pi} {\cal N}_{ik}  
\,
 2P_{kl}^{Q,(0)}(N){\cal D}_{lj}^{N}(\mu_0,\mu) \,.  
\eea 
% All ingredients needed to achieve  NLL resummation result can be found in~\cite{Cao:2023oef}.
\subsection{Transversely polarized QC}\label{sec:poltfr}
When the incoming nucleon is polarized, we can also probe the spin asymmetry by measuring the azimuthal angle.
The leading-twist quark contributions relevant to the
SSA are given by the following correlation matrix:
\bea\label{eq:fs} 
f_{q,{\rm QC}}(z,\theta,\phi,S_\perp) 
&=&  \int \frac{dy^-}{4\pi} e^{- i z P^+ \frac{y^-}{2} }  \langle P,S_\perp |
{\bar \chi}_n\left(\frac{y^-}{2}n^\mu\right) \frac{\gamma^+}{2} 
\hat{{\cal Q}}(\vec n)  \chi_n(0)
| P,S_\perp \rangle  \nn\\
&=& \frac{f_{q,{\rm QC}}(z,\theta)}{2\pi}+|S_\perp|\sin(\phi-\phi_S)f_{1T,q,{\rm QC}}(z,\theta)
\,, 
\eea 
where $f_{1T,q,{\rm QC}}(z,\theta)$ describe the unpolarized quark distributions in a transversely polarized nucleon. 
$(\phi - \phi_S)$ is the azimuthal angle between the detector and the
nucleon spin. 

Then we have
\bea
\Sigma(\theta,\phi,{S_T})=\Sigma_{UU}(\theta)+|S_\perp|\sin(\phi-\phi_S)\Sigma_{UT}
\eea
$\Sigma_{UT}$ can be factorized similarly as the unpolarized distribution in Eq.~(\ref{eq:fact-N}) with the replacement of $f_{q,{\rm QC}}$ by $f_{1T,q,{\rm QC}}(z,\theta)$.
The non-vanishing of the $\sin(\phi-\phi_S)$ dependent part arises from the same mechanism as the Sivers effect~\cite{Sivers:1989cc,Collins:2002kn}. The Sivers-like QC induces a $\sin(\phi-\phi_S)$ azimuthal asymmetry:
\begin{equation}
A^{Sivers} = \frac{\Sigma(S_T) - \Sigma(-S_T)}{\Sigma(S_T)+\Sigma(-S_T)}.
\end{equation}
The prediction of $A^{Sivers}$ relies on the non-perturbative input of the $f_{T,QC}$ which requires further study in the future. 

\section{TMD Region}
\label{sec:CFR}
\subsection{Unpolarized target}

We now consider the structure of the QC observable in the TMD region for an unpolarized target. Our analysis is based on the standard factorization of semi-inclusive DIS in the back-to-back limit in the Breit frame~\cite{Ji:2004wu}. In this region, the transverse momentum imbalance $q_T\ll Q$ is described by TMD beam, soft, and fragmentation functions, and the differential cross section can be written as
\bea
\label{eq:SIDIS_fac}
    \frac{d\sigma_{e+p\rightarrow e+a+X}}{dQ^2dx_Bd^2\mathbf{q}_Tdz}&=&\sigma_0
    H(Q^2,\mu;\mu_H)\sum_f
        \int  \frac{d^2\mathbf{b}}{(2\pi)^2} Q_f^2
        \nn \\
         &&\hspace{-10.ex}\times\exp[-i{\mathbf q}_T\cdot \mathbf{b}]B_{f/p}\left(x_B,b,\mu,\nu;\mu_B,\nu_B\right)\nn \\
        && \hspace{-10.ex}\times S(b,\mu,\nu;\mu_S,\nu_S)
       D_{a/f}(z,b,\mu,\nu;\mu_J,\nu_J)\,,
\eea
up to power corrections of ${\cal O}(q_T^2/Q^2)$. Here $
    \sigma_0 = \frac{2\pi\alpha^2}{Q^4}\left[1+\left(1-y\right)^2\right]$, $z$ denotes the longitudinal momentum fraction carried by the observed hadron, and an implicit sum over quark flavors is understood. The hard function $H$ encodes the short-distance scattering, while
$B_{f/p}$, $S$, and $D_{a/f}$ denote the TMD beam, soft, and fragmentation functions, respectively, $\mathbf{b}$ is the Fourier conjugate variable to $\mathbf{q}_T$ and $b\equiv |\mathbf{b}|$. In the following, we suppress the explicit dependence on the canonical scales $\mu_i$, $\nu_i$ for brevity.

In the back-to-back region, the charge-weighted angular distribution can be expressed in terms of the single-hadron semi-inclusive DIS (SIDIS) process. Introducing the observable $\frac{d\Sigma(\theta)}{d\theta}$, we write
\bea
\label{eq:qcsidis}
\frac{d\Sigma(\theta)}{d\theta}
=
\sum_a\int d^2\mathbf{q}_T\,dz\;
\frac{d\sigma_{e+p\rightarrow e+a+X}}{dQ^2\,dx_B\,d^2\mathbf{q}_T\,dz}\;
Q_a\,
\delta(\theta_{ap}-\theta)\,,
\eea
where $Q_a$ is the hadron charge and $\theta_{ap}$ is the angle between the observed hadron and the proton beam direction in the Breit frame. Using $\theta_{ap}\sim \pi-\frac{2|{\mathbf q}_{T}|}{Q}$, the factorized form becomes 
\bea
\label{eq:TMDfac}
    \frac{d\Sigma(\theta)}{d\theta}&=& \sigma_0
        H(Q^2,\mu) \sum_fQ_f^2
        \int  d^2{\mathbf q}_{T}\,  \frac{d^2\mathbf{b}}{(2\pi)^2} \exp[-i{\mathbf q}_T\cdot \mathbf{b}]\nn \\
         &\times&B_{f/p}\left(x_B,b,\mu,\nu\right)S(b,\mu,\nu)
       J_{f,\text{Q}}(b,\mu,\nu)\delta\left(\frac{2|{\mathbf q}_{T}|}{Q}-\bar{\theta}\right)\, ,
\eea 
where $\bar{\theta}\equiv\pi-\theta$, and $J_{f,\text{Q}}$ is the charge-charge correlation jet function~\cite{Monni:2025zyv} defined as the charge-weighted first moment of the TMDFF:
\bea
    J_{f,\text{Q}}(b, \mu,\nu)\equiv\sum_a\int_0^1 dz\, Q_aD_{a/f}(z,b,\mu,\nu)\,.
\eea 
In the perturbative regime $1\gg (\pi-\theta)\sim q_T/Q$ with $(\pi-\theta)Q\gg \Lambda_{\rm QCD}$, both the QC jet function and the TMD beam function admit an operator-product expansion onto collinear degrees of freedom. As a result, they can be written as convolutions of perturbatively calculable short-distance matching coefficients with the corresponding fragmentation functions(FF) $D_{i/f}\left(\frac{\omega}{z},\mu_0\right)$~\cite{Monni:2025zyv} and collinear PDFs: 
\bea
\label{eq:QCjetope}
    &&J^{\text{OPE}}_{f,\text{Q}}(b, \mu,\nu)=\sum_j\int_0^1 d\omega \,Q_j{\cal D}_{fj}\left(\omega,\frac{b}{\omega},\mu,\nu\right)\, ,
\\
&&B^{\text{OPE}}_{f/p}(x,b,\mu,\nu) =\sum_{i} \int_x^1 \frac{dz}{z}  {\cal I}_{fi}\left(b, \frac{x}{z},\mu,\nu\right) f_{i/p}(z,\mu)\, .
\eea
 ${\cal D}_{fj}$ and ${\cal I}_{fi}$ are perturbatively calculable Wilson coefficients and $f_{i/p}$ denotes the usual collinear PDFs. We have used the superscript OPE to denote that this is the leading contribution in the expansion. In addition, we have employed the sum rule 
 \bea
 Q_f=\sum_a\int_0^1dz \,Q_a D_{a/f}\left(z,\mu\right)\, ,
 \eea
which leads to the absence of FF in the first line of Eq.~(\ref{eq:QCjetope}).
 
The beam, jet, and soft functions can be evolved to the common scale $\mu$ from their natural scales at $\nu_B$, $\nu_J$, $\nu_S$, $\mu_B$, $\mu_J,$ and $\mu_S$, respectively:
\bea
\label{eq:BJSevolPos}
B_f(x,b,\mu,\nu;\mu_B,\nu_B) &=& U_B(\mu,\nu;\mu_B,\nu_B)
 B_f(x,b,\mu_B,\nu_B), \nn \\
J_{f,\rm{QC}}(b,\mu,\nu;\mu_J,\nu_J) &=& U_J(\mu,\nu;\mu_J,\nu_J) J_{f,\rm{QC}}(b,\mu_J,\nu_J),  \\
S(b,\mu,\nu;\mu_S,\nu_S) &=& U_S(\mu,\nu;\mu_S,\nu_S) S(b, \mu_S,\nu_S),\nn
\eea
where $U_B$, $U_J$, and $U_S$ are the position-space evolution factors for the beam, jet, and soft functions. Similarly, the hard function also has a multiplicative renormalization group evolution
\bea
\label{eq:hardfuncevol}
H(Q^2, \mu; \mu_H) &=& U_H (Q^2,\mu, \mu_H)H(Q^2, \mu_H)\,,
\eea
where $U_H (\xi^2,\mu, \mu_H)$ is the corresponding hard-function renormalization-group evolution factor. Combining these ingredients, the RG-improved factorization formula becomes
\bea
\label{eq:TMDfacRG}
    \frac{d\Sigma(\theta)}{d\theta}&=&
        H(Q^2,\mu_H)\sum_fQ_f^2
        \int  d^2{\mathbf q}_{T}\,  \frac{d^2\mathbf{b}}{(2\pi)^2} 
        \nn \\
        &&\hspace{-10.ex}\times \exp[-i{\mathbf q}_T\cdot \mathbf{b}]U_{tot}B_{f/p}\left(x_B, b,\mu_B,\nu_B\right)
        \nn \\
          &&\hspace{-10.ex}\times 
          S(b,\mu_S,\nu_S)
       J_{f,\text{Q}}(b,\mu_J,\nu_J)   
       \delta\left(\frac{2|{\mathbf q}_{T}|}{Q}-\bar{\theta}\right)\, ,
\eea 
where we have introduced the shorthand $U_{\rm tot}\equiv U_B\,U_J\,U_S\,U_H$.

The factorization  outlined above provides a systematic way to study the QC in the TMD region and allows for the resummation of large logarithms. It enables the calculation of precise theoretical predictions for the observable in high-energy scattering processes involving hadrons. 
\begin{table}[htb]   

\begin{center}
\begin{tabular}{|c|c|c|c|c|}
\hline   Accuracy & $H$, $J$, $S$, $B$ & $\gamma_{\text{cusp}}$ & $\gamma$ & $\beta$  \\ 
 \hline   NLL & Tree & 2 loop & 1 loop & 2 loop  \\    
\hline   NNLL & 1 loop & 3 loop & 2 loop & 3 loop  \\    
\hline   N$^3$LL & 2 loop & 4 loop & 3 loop & 4 loop  \\    
\hline   
\end{tabular}
\caption{Classification of the resummation accuracy in terms of the fixed-order expansions of boundary term, anomalous dimensions, and beta function in the TMD region}
\label{table:1}
\end{center}   
\end{table}
Table~\ref{table:1} lists the ingredients required for resummation up to N$^3$LL. The hard function is known at $O(\alpha_s^2)$~\cite{Idilbi_2006,Becher:2006mr}. The soft function has been calculated at $O(\alpha_s^2)$ in~\cite{Moult:2018jzp,Li:2016ctv}. The QC jet function is available up to $O(\alpha_s^3)$~\cite{Monni:2025zyv}. The beam function has been calculated up to $O(\alpha_s^3)$~\cite{Gehrmann:2012ze,Gehrmann_2014,L_bbert_2016,Echevarria_2016,Luo:2019bmw,Luo:2019hmp,Luo:2020epw,Luo:2019szz,Luo_2021}. Finally, the analytic expression for the four loop cusp anomalous dimension, needed to solve the renormalization group evolution equations at N$^3$LL was obtained in~\cite{Henn:2019swt,Moult:2022xzt}.

When $b$ becomes large and thus $\mu_b\lesssim \Lambda_{\rm QCD}$, the TMD evolution enters the non-perturbative region. We follow the usual $b^*$-prescription~\cite{Collins:1984kg} that introduces a cut-off value $b_{\rm max}$ and allows for a smooth transition from the perturbative to the non-perturbative region,
\bea
b^* = b/\sqrt{1+b^2/b_{\rm max}^2}\,,
\eea
with $b_{\rm max} = 1.5$ GeV$^{-1}$. 

\subsection{Transversely Polarized Target}

The differential SIDIS cross section on a transversely polarized nucleon target can be written 
as~\cite{Kang:2012xf,Bacchetta:2006tn,Anselmino:2008sga}
\bea
\frac{d\sigma_{e+p(S_\perp)\rightarrow e+a+X}}{dx_B dQ^2 dz d^2\bf{q}_{T}}
&= 
\left[F_{UU} + |S_\perp|\sin(\phi_h-\phi_s)\,
F_{UT}^{\sin\left(\phi_h -\phi_s\right)}\right],
  \label{eq:aut}
\eea
where $\phi_s$ and $\phi_h$ are the azimuthal angles for the nucleon spin and the transverse momentum 
of the outgoing hadron, respectively.
$F_{UU}$ is the unpolarized distribution defined in Eq.~(\ref{eq:SIDIS_fac}) divided by $2\pi$ and $F_{UT}^{\sin(\phi_h-\phi_s)}$ is the transverse spin-dependent structure function:
\bea
 F_{UT}^{\sin(\phi_h-\phi_s)}(x_B, z,|{\bf q}_{T}|, Q) &= & \sigma_0 H(Q;\mu) \sum_f Q_f^2 \int_0^{\infty} \frac{b^2\, db}{4\pi}J_1\left(\frac{b |{\bf q}_{T}|}{z} \right) \nn\\
    & \times& f_{1 T, f/p}(x_B,b,\mu,\nu)S(b,\mu,\nu)D_{h/f}(z,b,\mu,\nu)  \, .
\label{fut}
\eea
Here, $f_{1T,q/p}(x_B, b;\mu,\nu)$ is the SIDIS Sivers function and $J_1$ is the Bessel function of the first order.

At small $b$ where $1/b \gg \Lambda_{\rm QCD}$, one can perform an operator product expansion (OPE) of the the SIDIS Sivers function onto their collinear counterparts~\cite{Echevarria:2020hpy,Ji:2006ub,Ji:2006vf,Koike:2007dg,Kang:2011mr,Sun:2013hua,Dai:2014ala,Scimemi:2019gge}, we follow the convention in~\cite{Echevarria:2020hpy} and get:
\begin{align}
    f_{1T,q/p}^{ \text{OPE}} (x, b,\mu,\nu) =U_B(\mu,\nu;\mu_B,\nu_B) \int_{x}^{1} \frac{d\hat{x}_1}{\hat{x}_1}\frac{d \hat{x}_2}{\hat{x}_2} \bar{C}_{q\leftarrow i}(x/\hat{x}_1,x/\hat{x}_2,b,\mu_B,\nu_B) \, 
T_{F\, i/p}(\hat{x}_1,\hat{x}_2,\mu_B) \,,
\end{align}
As the Sivers function obeys the same evolution equations as the unpolarized TMD PDF, the evolution kernel $U_B$ is identical to that of the TMD PDF. $T_{F\, i/p}(x_1, x_2, \mu)$ is the Qiu-Sterman function~\cite{Echevarria:2020hpy,Sun:2013hua,Scimemi:2019gge}, and 
\bea
\bar{C}_{q\leftarrow i}\left(x_1,x_2,b;\mu_{b^*},\nu=Q\right)  &= &
\delta_{qi}\, \delta(1-x_1)\, \delta(1-x_2) \nn\\
&-& \frac{\alpha_s}{2\pi}\delta_{qi}
\frac{1}{2N_C}\delta(1-x_2/x_1)(1-x_1)
\,,
\eea
where $\mu_{b^*}\equiv\frac{2e^{-\gamma_E}}{b^*}$. To get $T_{F\, i/p}(\hat{x}_1,\hat{x}_2,\mu_{b^*})$, we parameterization $T_{F\, i/p}(\hat{x},\hat{x},\mu_0)$ to be proportional to the unpolarized PDF $f_{i/p}(x, \mu_0)$ at the initial scale $\mu_0=\sqrt{1.9}$ GeV~\cite{Echevarria:2014xaa, Echevarria:2020hpy}: 
\begin{align}
    T_{F\, q/p}(x,x,\mu_0) = N_q\frac{\left( \alpha_q+\beta_q \right)^{\left( \alpha_q+\beta_q \right)}}{\alpha_q^{\alpha_q} \beta_q^{\beta_q}}x^{\alpha_q}(1-x)^{\beta_q}\,
f_{q/p}(x,\mu_0)\,.
\end{align}

In order to obtain a numerical result, evolution of the Qiu-Sterman function must be performed from $\mu_0$ to the natural scale $\mu_{b^*}$ with~\cite{Kang:2012em}
\begin{align}
    \frac{\partial T_{F\, q/p}(x,x;\mu)}{\partial \ln \mu^2}
    = \frac{\alpha_s(\mu^2)}{2\pi} \int dz \left[P^{(0)}_{qq}(z)-N_C\delta(1-z)\right] 
    T_{F\, q/p}\left(x/z,x/z;\mu\right)\,.
\end{align}
Following the procedure outlined in the last section, we can get
 \bea
\label{eq:qcsidispol}
\frac{d\Sigma(\theta,\phi_h)}{d\theta}
=
\frac{d\Sigma_{UU}(\theta)}{d\theta}+\sin(\phi_h-\phi_s)\frac{d\Sigma^{\sin(\phi_h-\phi_s)}_{UT}(\theta)}{d\theta}\, ,
\eea
where
\bea
\label{eq:TMDpolfac}
    \frac{d\Sigma^{\sin(\phi_h-\phi_s)}_{UT}(\theta)}{d\theta}&=& \sigma_0
        H(Q^2,\mu) \sum_fQ_f^2 \int {\bf q}_{T} d{\bf q}_{T}\int_0^{\infty} \frac{b^2\, db}{4\pi}J_1\left({b |{\bf q}_{T}|} \right)\nn \\
         &\times&f_{1 T, f/p}(x_B,b,\mu,\nu)S(b,\mu,\nu)
       J_{f,\text{Q}}(b,\mu,\nu)\delta\left(\frac{2|{\mathbf q}_{T}|}{Q}-\bar{\theta}\right)\, ,
\eea 
We can define the asymmetry as the ratio of the above structure functions:
\bea
    \frac{dA^{Sivers}}{d\theta}&=& \frac{{d\Sigma^{\sin(\phi_h-\phi_s)}_{UT}(\theta)}/{d\theta}}{{d\Sigma_{UU}(\theta)}/{d\theta}}=\frac{{d\Sigma^{\sin(\phi_h-\phi_s)}_{UT}(\theta)}/{d\theta}}{{d\Sigma(\theta)}/{(2\pi d\theta)}}\,.
\eea 

\subsection{Non perturbative effect}
 A key feature of Eq.~(\ref{eq:TMDfacRG}) is that the same universal back-to-back soft function enters as in standard TMD observables. This allows us to incorporate the non-perturbative contributions using the conventional TMD modeling strategy. For the hadronization model of the QC jet
function we can assume a generic multiplicative ansatz. The
final result reads,
\begin{align}
    \sqrt{S}J_{f,\text{Q}}(b,\mu_0,\nu_0)&=\sqrt{S_{\text{pert.}}}J^{\text{OPE}}_{f,\text{Q}}(b,\mu_0,\nu_0)j_{f,Q}^{\rm NP}(b)\,,
\end{align}
where $S_{\text{pert.}}$ denotes the perturbative soft function, while $j_Q^{NP}(b)$ parametrize non-perturbative effects in the jet sector. Since this is the same
 function that appears in the QC observable in~\cite{Monni:2025zyv} one
can extract the model function $j_Q^{NP}(b)$ by fitting to experimental data from $e^+e^-$ process. 
Alternatively, we adopt the strategy used in the EC study~\cite{Li:2021txc}, and relate the model function  $j_Q^{NP}(b)$ to the standard TMDFFs as follows,
\bea
\label{modj}
j_{f,Q}^{NP}(b)J^{\text{OPE}}_{f,\text{QC}}(b, \mu_0,\nu_0)=\sum_{a,j}\int_0^1 \frac{d\omega}{\omega}\int_\omega^1\frac{dz}{z}\, Q_a{\cal D}_{fj}\left(\frac{ b}{z},z,\mu_0,\nu_0\right) D_{a/j}\left(\frac{\omega}{z},\mu_0\right)d^{\rm NP}(\omega,b)\,.
\eea
where $D_{i/f}\left(z,\mu_0\right)$ is the FF and the non-perturbative model function $d_{NP}(\omega,b)$ is defined in the context of TMDFFs~\cite{Echevarria:2020hpy},
\bea
    \sqrt{S}D_{a/f}(z,b,\mu_0,\nu_0)=\sqrt{S_{\text{pert.}}}D^{\text{OPE}}_{a/f}(z,b,\mu_0,\nu_0)d^{\rm NP}(z,b)\,,
\eea
Therefore, given a specific model for TMD fragmentation we
can explicitly evaluate the model function $j_Q^{NP}(b)$.
Recent extractions of TMD fragmentation functions are available in~\cite{Sun:2014dqm,Bacchetta:2017gcc,Scimemi:2019cmh}. The non-perturbative components of the TMD PDFs and the Sivers function can be modeled following the same strategy commonly adopted in TMD analyses of SIDIS and Drell–Yan processes~\cite{Echevarria:2020hpy},
\bea
\label{eq:model-PDF}
   &&\sqrt{S} B_{i/P}(x,b;\mu_0,\nu_0) = \sqrt{S^{\text{pert.}}} B_{i/P}^{\text{OPE}}(x,b;\mu_0,\nu_0)
     f_{i/P}^{\text{NP}}(b)\, ,\nn \\
     &&\sqrt{S} f_{1T,q/p} (x, b,\mu_0,\nu_0) = \sqrt{S^{\text{pert.}}} f_{1T,q/p}^{\,\text{OPE}}(x,b;\mu_0,\nu_0)
     f_{i/P,T}^{\text{NP}}(b)\;.
\eea
We combine all non-perturbative ingredients and write them as 
\bea
F_{i/P}^{\rm NP}(b)&\equiv& f_{i/P}^{\text{NP}}(b)j_{i,Q}^{NP}(b)\,,\nn
\\
F_{i/P,T}^{\rm NP}(b)&\equiv& f_{i/P,T}^{\text{NP}}(b)j_{i,Q}^{NP}(b)\,.
\eea
Finally the distribution can be written as 
\bea
\label{eq:TMDfacRGf}
    \frac{d\Sigma(\theta)}{d\theta}&=&
        H(Q^2,\mu_H)
        \int  d^2{\mathbf q}_{T}\,  \frac{d^2\mathbf{b}}{(2\pi)^2} 
        \nn \\
        &&\hspace{-10.ex}\times \exp[-i{\mathbf q}_T\cdot \mathbf{b}]U_{tot}B^{\rm OPE}_{f/p}\left(x_B, b,\mu_B,\nu_B\right)
        \nn \\
          &&\hspace{-10.ex}\times 
          S_{\rm pert}(b,\mu_S,\nu_S)
       J^{\rm OPE}_{f,\text{Q}}(b,\mu_J,\nu_J)   F_{i/P}^{\rm NP}(b)
       \delta\left(\frac{2|{\mathbf q}_{T}|}{Q}-\bar{\theta}\right)\, ,
\eea 
and
\bea
\label{eq:TMDpolfacf}
    \frac{d\Sigma^{\sin(\phi_h-\phi_s)}_{UT}(\theta)}{d\theta}&=& \sigma_0
        H(Q^2,\mu) \sum_fQ_f^2 \int  {\bf q}_{T}d{\bf q}_{T}\int_0^{\infty} \frac{b^2\, db}{4\pi}J_1\left({b |{\bf q}_{T}|} \right)\nn \\
         &\times&f^{\rm OPE}_{1 T, f/p}(x_B,b,\mu,\nu)S_{\rm pert}(b,\mu,\nu)
       J^{\rm OPE}_{f,\text{QC}}(b,\mu,\nu)F_{i/P,T}^{\rm NP}(b)\delta\left(\frac{2|{\mathbf q}_{T}|}{Q}-\bar{\theta}\right)\, ,
\eea
Following~\cite{Echevarria:2020hpy,Sun:2014dqm}, we adopt the nonperturbative model
\bea
d_{NP}(z,b)&=&\exp\left(-\frac{g_2}{2}\ln\frac{Q}{Q_0}\ln\frac{b}{b^*}-g_1^D b^2\right)\nn\\
f_{i/P}^{\text{NP}}(b)&=&\exp\left(-\frac{g_2}{2}\ln\frac{Q}{Q_0}\ln\frac{b}{b^*}-g_1^fb^2\right)\nn \\
f_{i/P,T}^{\text{NP}}(b)&=&\exp\left(-\frac{g_2}{2}\ln\frac{Q}{Q_0}-g_1^Tb^2\right)\,.
\eea
with
\bea
    g_2 = 0.84\,, \quad Q_0 = \sqrt{2.4}\, \text{GeV}\,, \quad g_1^f = 0.106\,, \quad g_1^D = 0.042\,, \quad g_1^T=0.18 .
\eea
For the numerical implementation, we need a model for the jet function in Eq.~(\ref{modj}).
For simplicity we only use the matching coefficients at LO and choose the following simple parametrization:
\begin{equation}
j_{i,Q}^{\rm NP}(b) = \exp\left(-\frac{g_2}{2}\ln\frac{Q}{Q_0}-g_i^jb^2\right)\, .
\end{equation}
    where $g_i^j$ are free parameters. We perform a fit with the NPC23 collinear fragmentation functions for charged hadrons~\cite{Gao:2024dbv} in the region $2<b<10$ GeV$^{-1}$ for $i\in\{u,d,\bar{u},\bar{d}\}$ and obtain $g_i^j=0.14$. We further shift the input variation by $\pm50\%$, finding that the resulting effect is negligible.
We take this result as the starting point for our numerical analysis.

\section{Numerical Results }
\label{sec:num}

In this section, we study the QC observable at $\sqrt{s}=140~\mathrm{GeV}$ with $400 < Q^2 < 500~\mathrm{GeV}^2$, relevant for a future EIC. For all calculations, we use the {\tt NNPDF40\_nnlo\_as\_01180} PDF sets~\cite{NNPDF:2021njg}.

\begin{figure}[!htbp]
  \centering

    \centering
    \includegraphics[width=0.45\textwidth]{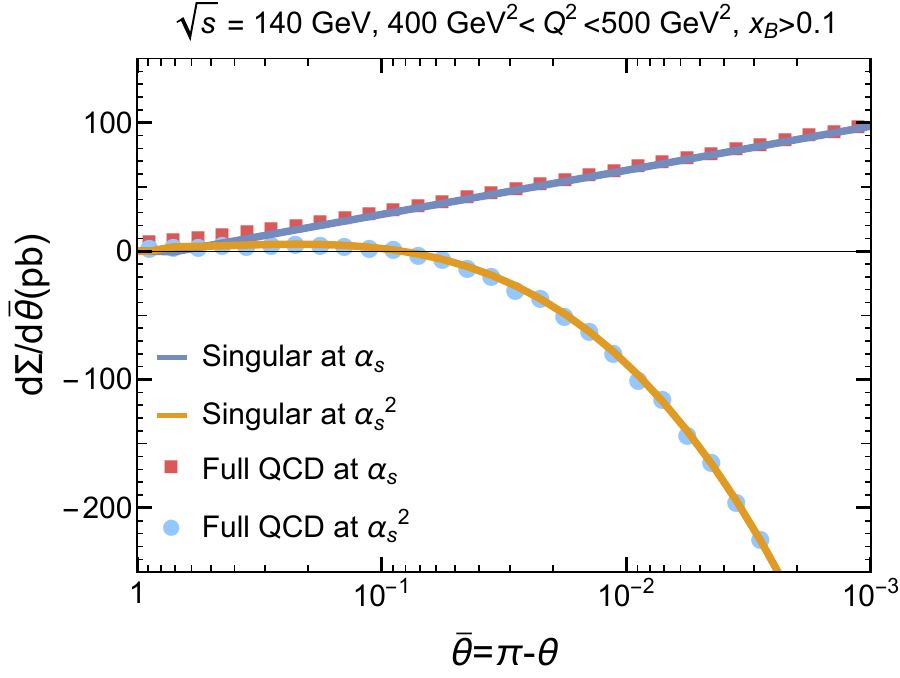}
    \includegraphics[width=0.45\textwidth]{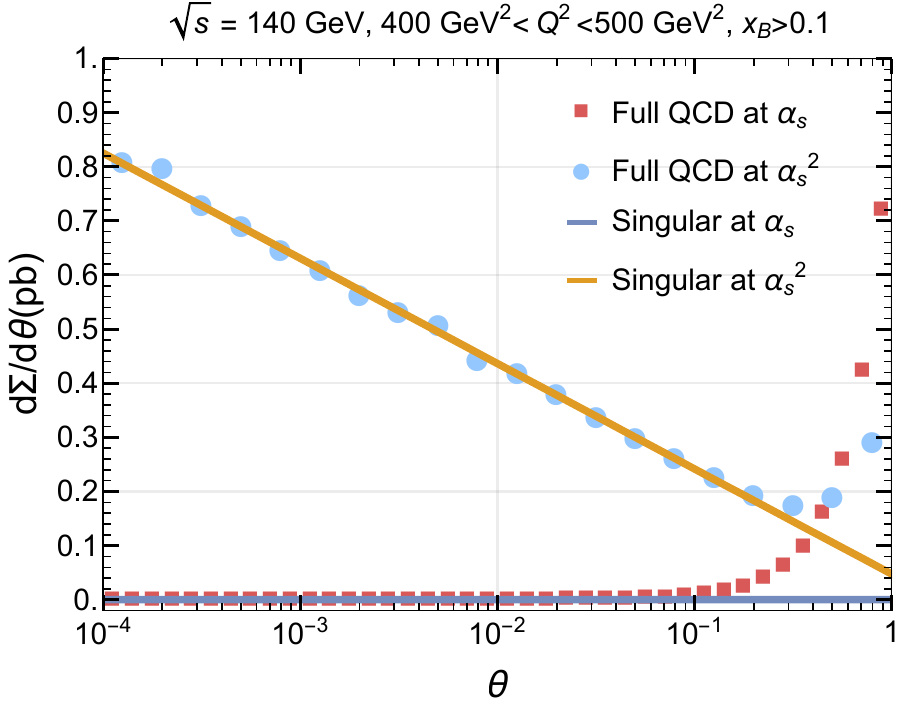}

    \caption{Comparison between the $\ln\theta$ leading singular contributions
with the full fixed-order calculations. The left panel shows the comparison for $\theta>\frac{\pi}{2}$ while the right panel shows the comparison for $\theta<\frac{\pi}{2}$. }
  \label{fg:sing}
  % \vspace{-5.ex}
\end{figure}

We first validate the factorization framework by comparing the leading singular terms in $\ln(\theta)$  predicted by SCET with the full QCD prediction at $\alpha_s$ and $\alpha_s^2$. We focus on the distribution $d\Sigma/dy$, where $y\equiv\ln\tan(\theta/2)$.
In the TFR and TMD regions, the logarithmically-enhanced $\ln(\theta)$ contributions dominate the cross section, and the singular approximation should reproduce the complete fixed-order prediction. The comparison is presented in Fig.~\ref{fg:sing}, with all scales chosen as $\mu=Q$. The full fixed-order results are obtained numerically using {\tt distress}~\cite{Abelof:2016pby}. In both the TFR and TMD regions we find excellent agreement between the singular terms derived from the factorization theorem and the complete QCD calculation, providing a non-trivial validation of the factorization formula. 

In both the TMD and TFR regions, large logarithmic contributions can significantly enhance higher-order corrections and degrade the reliability of the fixed-order perturbative expansion. The resummation of these logarithms to all orders in the strong coupling is essential for reliable predictions. In the TMD regime, the resummed cross section is obtained by evolving the hard, soft, beam, and jet functions appearing in Eq.~(\ref{eq:TMDfac}) from their natural scales to common renormalization and rapidity scales, denoted by $\mu$ and $\nu$. In our numerical implementation, we adopt the canonical scale choices
\bea
\label{eq:scales}
\mu=\mu_H=\nu=\nu_J=\nu_B=Q, 
\qquad 
\mu_J=\mu_S=\mu_B=\frac{2e^{-\gamma_E}}{b^*}\,.
\eea

\begin{figure}[!htbp]
    \begin{minipage}{\linewidth}
    \includegraphics[width=0.6\textwidth]{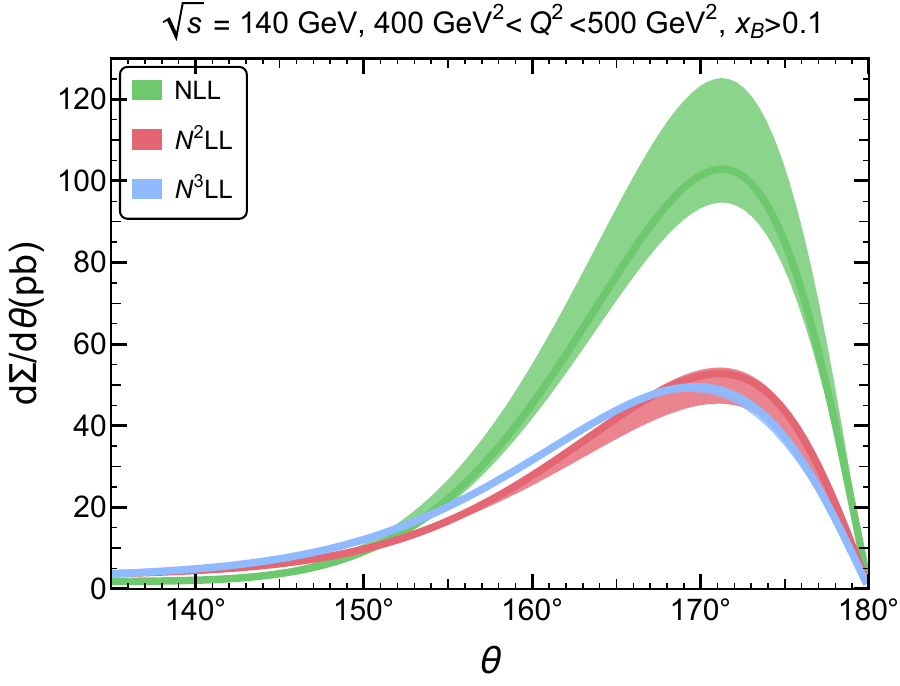}
    \end{minipage}
    \caption{Resummed $y$ distributions for the QC in the TMD region.}
    \label{fg:error}
\end{figure}

Fig.~\ref{fg:error} presents the resummed distributions in the TMD region.The theoretical uncertainty is estimated by independently varying the scales $\mu=\mu_H$, and $\nu$ up and down by a factor of two around their central values. We observe sizable corrections when going from NLL to N$^2$LL, while the N$^2$LL and N$^3$LL results are in much better agreement. The scale uncertainties are reduced at N$^3$LL compared to lower-order predictions.

\begin{figure}[!htbp]
    \includegraphics[width=0.45\textwidth]{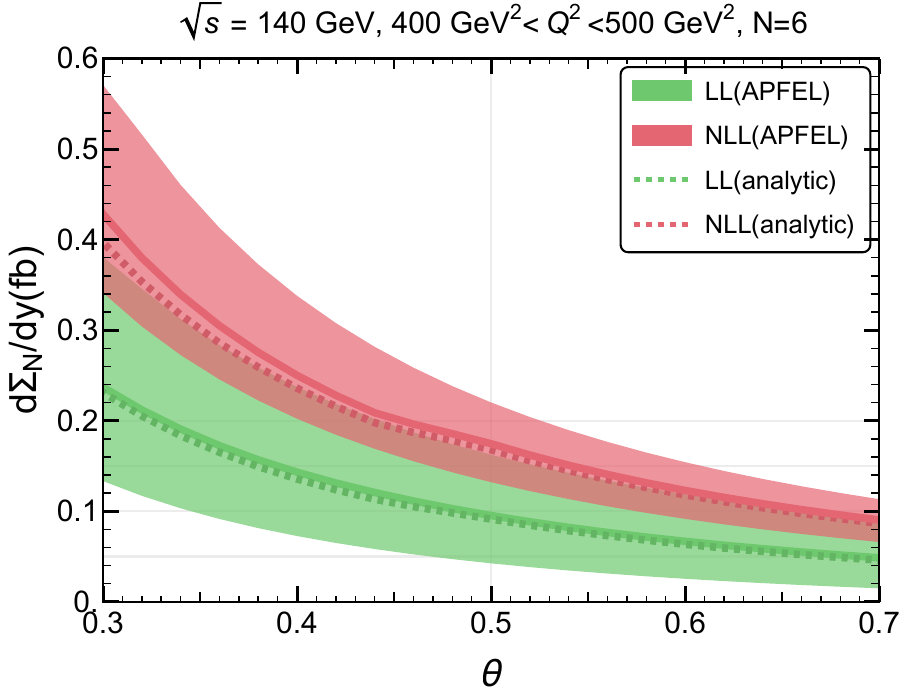}
    \includegraphics[width=0.45\textwidth]{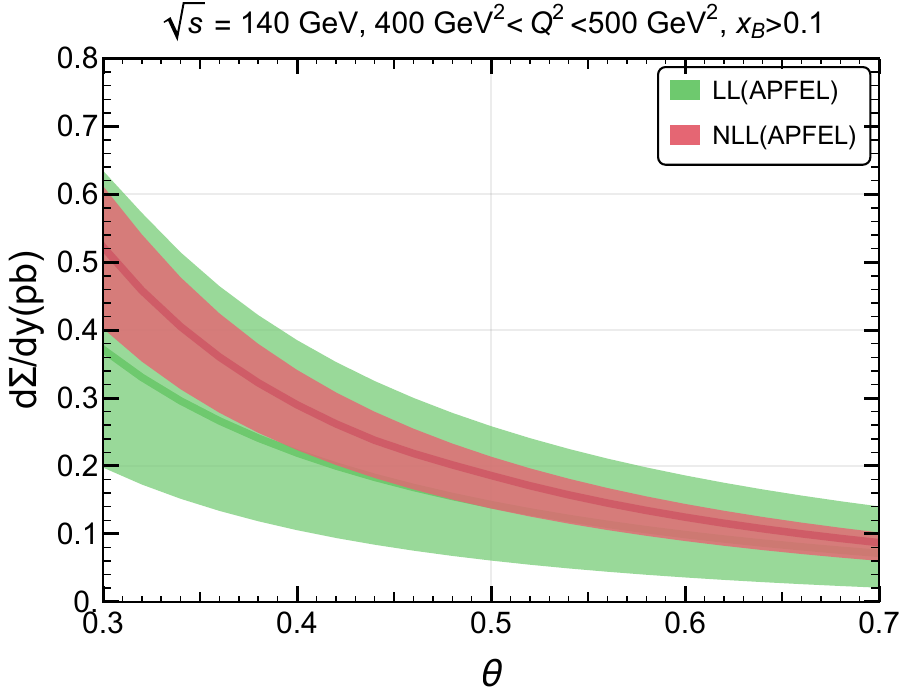}
    \caption{Resummed $y$ distributions for the QC within the TFR shown in Mellin space in the left panel and in $z$-space in the right panel. }
    \label{fg:errorTFR}
\end{figure}

Fig.~\ref{fg:errorTFR} presents the resummed distributions in the TFR. We choose $\mu_h=\mu$, allowing us to evaluate the resummed cross section by evolving the QC from $\mu_0$ to $\mu$. In this case, we select the canonical resummation scales as follows:
\bea
\label{eq:scalesTFR}
\mu=Q, \qquad \mu_0=\frac{Q\theta}{2}\,.
\eea
The scale uncertainties are evaluated by varying $\mu_0$ up and down by a factor of 2 independently. 

The left panel of Fig.~\ref{fg:errorTFR} illustrates the resummed distributions in the TFR region in Mellin space for $N=6$. We show this plot to confirm our numerical treatment through the comparison to this analytic result. For $\theta<0.3$, $\mu_0/2$ becomes comparable to $\Lambda_{\text{QCD}}$, and the perturbative calculation is no longer reliable in this regime. The band shows the prediction from {\tt APFEL}~\cite{Bertone:2013vaa,Bertone:2017gds}, while the dashed line corresponds to the analytic Mellin-space result obtained from Eq.~(\ref{eq:fNLLevo}) at the central scale. This provides a cross-check of our numerical implementation. Owing to the factor $x_B^{N-1}$, the cross section is significantly suppressed compared to the full $z$-space result, shown in the right panel of Fig.~\ref{fg:errorTFR}. Although the resummed result in the TFR region is significantly lower than that in the TMD region, the expected integrated luminosity at the EIC, more than $10^{4}\,\text{pb}^{-1}/\text{year}$~\cite{AbdulKhalek:2021gbh}, suggests that a sizable number of events can still be observed in this kinematic regime.

\begin{figure}[!htbp]
    \begin{minipage}{\linewidth}
    \includegraphics[width=0.6\textwidth]{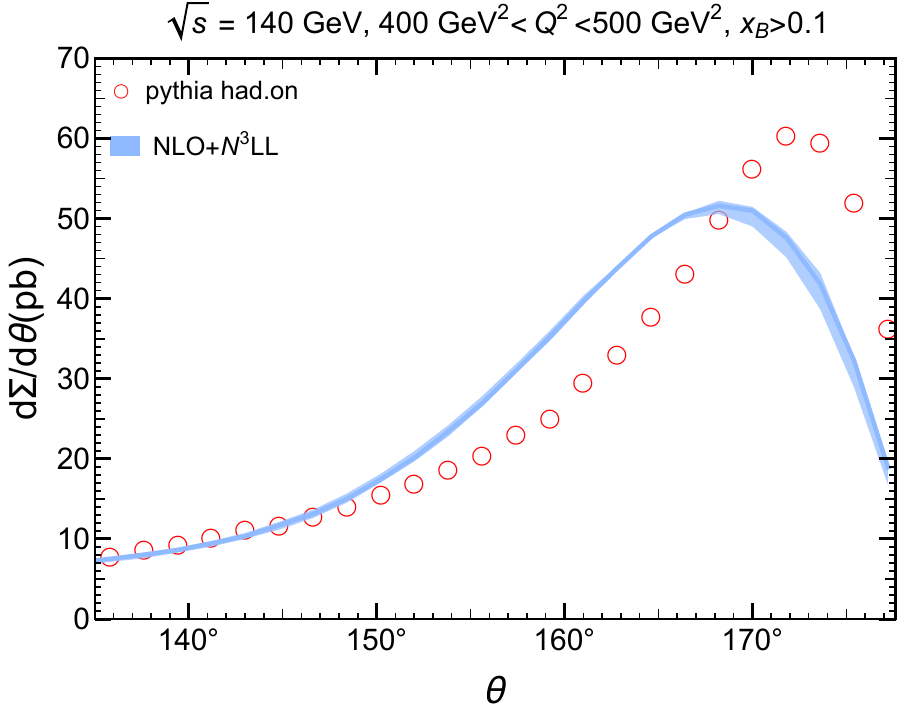}
    \end{minipage}
    \caption{Comparison of the distributions between the N$^3$LL+NLO prediction, the QCD prediction and \textsc{PYTHIA} simulations. }
    \label{fg:cfrpythia}
\end{figure}

The prediction of N$^3$LL+NLO in the TMD region is presented in  Fig.~\ref{fg:cfrpythia}. The N$^3$LL+NLO is achieved with
\bea
\label{eq:match}
\frac{d\Sigma(\theta)}{d{\theta}}&=&\frac{d\Sigma(\theta)}{d{\theta}}\Bigg|_{\text{resum}}+\frac{d\Sigma(\theta)}{d{\theta}}\Bigg|_{\text{QCD}}
-\frac{d\Sigma(\theta)}{d{\theta}}\Bigg|_{\text{resum exp. to NLO}}
\eea
The resummed result accounting for the effects of large logarithms to all orders correctly captures the behavior of PYTHIA in this region. 

\begin{figure}[!htbp]
    \begin{minipage}{\linewidth}
    \includegraphics[width=0.6\textwidth]{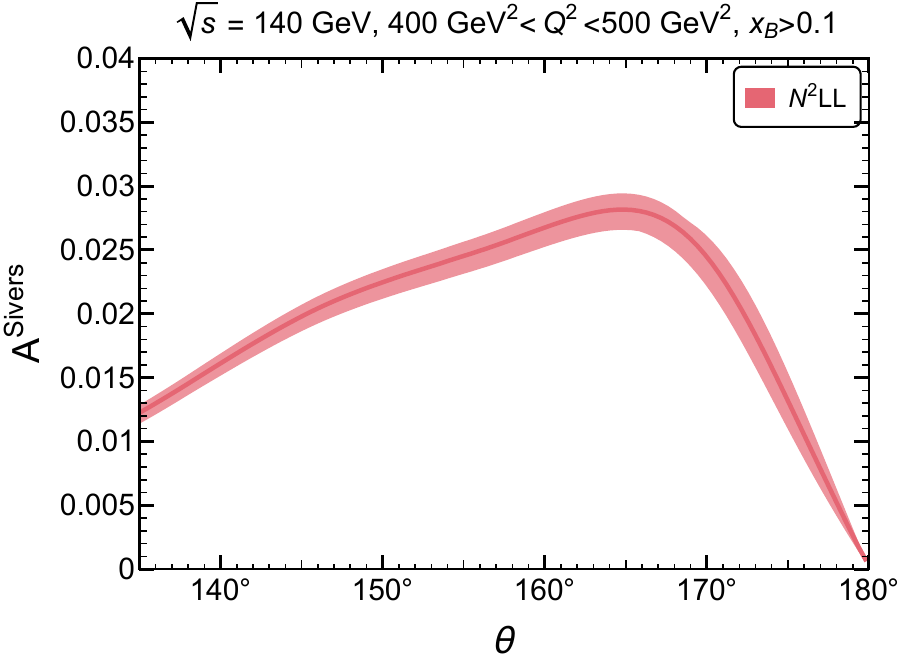}
    \end{minipage}
    \caption{Resummed y distributions for the Sivers asymmetry. The error band corresponds to the uncertainties in the parameters of the Sivers function in~\cite{Echevarria:2020hpy}}
    \label{fg:siver}
\end{figure}
In Fig.~\ref{fg:siver}, we present the results for the Sivers asymmetry using the parametrization in~\cite{Echevarria:2020hpy}. In this plot, in order to be consistent with the PDF set used in their parameter extraction, we employ the {\tt HERAPDF20\_NLO\_ALPHAS\_118} PDF set~\cite{H1:2015ubc}. The magnitude of the asymmetry ranges within a few percent and indicates that this asymmetry is a reasonable measurement for constraining the Sivers function.

\section{Conclusion}
\label{sec:con}
In this work, we have proposed a novel definition of the one-point charge correlator adapted to the Breit frame in DIS. This new observable provides a powerful and clean laboratory to investigate the structure of the nucleon. Utilizing SCET, we have systematically established the factorization theorems and provided precise theoretical predictions in both the TMD region and the TFR. Specifically, in the TFR, similar to the recently proposed nucleon energy correlators~\cite{Liu:2022wop}, the measurement of this charge correlator is remarkably clean, requiring no additional non-perturbative inputs beyond the nucleon charge correlator itself. Furthermore, due to its charge-sensitive nature, this observable exhibits significant potential for flavor tagging in future phenomenological analyses.

Looking forward, this novel framework opens up several exciting directions for future research. First, following recent developments in studying the odderon in the small-$x$ region~\cite{Bhattacharya:2025bqa, Mantysaari:2025mht}, it would be highly interesting to explore the odderon signatures within our charge correlator framework, which could be accessed in both the TMD region and the TFR. Second, extending the current unpolarized and Sivers asymmetry studies to investigate the comprehensive spin-dependent structure of the nucleon, along the lines of Ref.~\cite{Gao:2025cwy,Kang:2023big}, represents another compelling avenue. 

From an experimental perspective, this observable is relying solely on charged-particle tracks. It takes advantage of excellent angular resolution and does not need energy measurements, hadron identification, or jet-finding algorithms. This clean feature makes it a very promising tool for the EIC, where it can offer a unique way to study the structure and spin-dependent behavior of hadrons.

\begin{acknowledgments}
We thank Xiaohui Liu and Win Lin for useful discussions. H.~C. and F.~P. are supported by the U.S. Department of Energy, Office of High Energy Physics, under contract No. DE-SC0010143. H.~C. is partially supported by a CFNS  Joint Postdoctoral Fellowship. This research was supported in part through the computational resources and staff contributions provided for the Quest high performance computing facility at Northwestern University which is jointly supported by the Office of the Provost, the Office for Research, and Northwestern University Information Technology. 
\end{acknowledgments}

\begin{widetext} 
\appendix 

\section{solving the RG-evolution}\label{sec:derive}

In this section, we solve Eq.~(\ref{eq:evo2}), which can be written as
\bea\label{eq:inteform} 
f_{{\rm QC}}(N,\ln\frac{Q\theta}{u \mu})  
=  
f_{{\rm QC}}(N,\ln\frac{Q\theta}{u \mu_0})+ 
\int^\mu_{\mu_0} d\ln {\mu'}^2
  \int d\xi \xi^{N-1} 
P\left(\xi\right)
 f_{{\rm QC}}(N,\ln\frac{Q\theta}{\xi \, u \mu'}) 
 \,.
\eea 
For simplicity, we have suppressed the subscripts. The product of the $P$'s should be treated as the matrix product.

We write the ansatz solution to Eq.~(\ref{eq:inteform}) as 
\bea 
f_{{\rm QC}}(N,\ln\frac{Q\theta}{u \mu}) 
= D(\mu,\mu_0)f_{{\rm QC}}(N,\ln\frac{Q\theta}{u \mu_0}) + R(\mu,\mu_0)\,,
%+ N(\mu,\mu_0,u)
\eea 
where $D$ and $R$ are to be determined and satisfy $D(\mu_0,\mu_0) = 1$ and $R(\mu_0,\mu_0)=0$. 

We plug the ansatz back into Eq.~(\ref{eq:inteform}), to find  
\bea\label{eq:start}
D(\mu,\mu_0)f_{{\rm QC}}(N,\ln\frac{Q\theta}{u \mu_0})
+ R(\mu,\mu_0)
&=& f_{{\rm QC}}(N,\ln\frac{Q\theta}{u \mu_0}) 
+\int_{\mu_0}^\mu d\ln{\mu'}^2 P(N,\mu') R(\mu',\mu_0)
\nn \\
&+ & 
\int^\mu_{\mu_0} d\ln {\mu'}^2
  \int d\xi \xi^{N-1} 
P\left(\xi\right) \, D(\mu',\mu_0)f_{{\rm QC}}(N,\ln\frac{Q\theta}{\xi u \mu_0})\,. 
\eea 
To realize the NLL resummation, we use the NLO result as the initial input at $\mu_0$, and manipulate Eq.~(\ref{eq:start}) as
\bea
D(\mu,\mu_0)f_{{\rm QC}}(N,\ln\frac{Q\theta}{u \mu_0})
+R(\mu,\mu_0) 
&=& f_{{\rm QC}}(N,\ln\frac{Q\theta}{u \mu_0}) \nn\\
&&\hspace{-15.ex} + \int^\mu_{\mu_0} d\ln {\mu'}^2
P\left(N,\mu'\right) \, \left[ D(\mu',\mu_0)f_{{\rm QC}}(N,\ln\frac{Q\theta}{ u \mu_0}) 
+ R(\mu',\mu_0)
\right] 
\nn \\
&&\hspace{-15.ex} - \frac{\alpha_s(\mu_0)}{2\pi}
\int^\mu_{\mu_0} d\ln {\mu'}^2 
\tilde{P}\left(N,\mu'\right)  
D(\mu',\mu_0)
[2P^{Q}(N)]f(N,\mu_0)
\,,
\eea 
where we have used the property that at NLO, the initial condition satisfies
\bea\label{eq:lnu} 
f_{\rm QC}(N,\ln\frac{Q\theta}{\xi u \mu_0}) 
= f_{\rm QC}(N,\ln\frac{Q\theta}{ u \mu_0}) 
-\frac{\alpha_s(\mu_0)}{2\pi}
\ln \xi 
[2P^{Q}(N,\mu_0)]f(N,\mu_0)\,. 
\eea
and applied the definition 
\bea 
\tilde{P}(N) = \int_0^1 d\xi \xi^{N-1} P(\xi) \ln\xi \,. 
\eea 

Now we repeat the above procedure, to replace the $D(\mu',\mu_0)f_{\rm QC}(N,\ln\frac{Q\theta}{\xi u \mu_0})+R(\mu',\mu_0)$ using Eq.~(\ref{eq:start}) to find 
\bea
D(\mu,\mu_0)f_{{\rm QC}}(N,\ln\frac{Q\theta}{u \mu_0})
+ R(\mu,\mu_0)
&=& f_{{\rm QC}}(N,\ln\frac{Q\theta}{u \mu_0})  
+ \int^\mu_{\mu_0} d\ln {\mu'}^2
P\left(N,\mu'\right) \, f_{{\rm QC}}(N,\ln\frac{Q\theta}{ u \mu_0})
\nn \\
& &  \hspace{-20.ex}
+ \int_{\mu_0}^\mu d\ln{\mu'}^2 P(N,\mu') \int_{\mu_0}^{\mu'} d\ln{\mu''}^2 P(N,\mu'')R(\mu'',\mu_0)
\nn \\
&&\hspace{-20.ex}
+ \int^\mu_{\mu_0} d\ln {\mu'}^2
P\left(N,\mu'\right) \,
\int^{\mu'}_{\mu_0} d\ln {\mu''}^2\, 
P\left(N,\mu''\right) 
D(\mu'',\mu_0)f_{{\rm QC}}(N,\ln\frac{Q\theta}{  u \mu_0})
\nn \\
&&\hspace{-20.ex}
- 
\frac{\alpha_s(\mu_0)}{2\pi}
\int^\mu_{\mu_0} d\ln {\mu'}^2
P\left(N,\mu'\right) \,
\int^{\mu'}_{\mu_0} d\ln {\mu''}^2
\tilde{P}\left(N,\mu''\right) 
D(\mu'',\mu_0)[2P^{Q}(N)]f(N,\mu_0)
\nn \\
&&\hspace{-20.ex}
-\frac{\alpha_s(\mu_0)}{2\pi}
\int^\mu_{\mu_0} d\ln {\mu'}^2 
\tilde{P}\left(N,\mu'\right)  
D(\mu',\mu_0)
[2P^{Q}(N)]f(N,\mu_0) \,, 
\eea 
which can be organized as 
\bea
&&D(\mu,\mu_0)f_{{\rm QC}}(N,\ln\frac{Q\theta}{u \mu_0}) 
+ R(\mu,\mu_0)
\nn \\ 
&=& \int^\mu_{\mu_0} d\ln {\mu'}^2
P\left(N,\mu'\right) \, 
\int^{\mu'}_{\mu_0} d\ln {\mu''}^2\, 
P\left(N,\mu''\right) 
\left[ D(\mu'',\mu_0)
f_{{\rm QC}}(N,\ln\frac{Q\theta}{u \mu_0})
+R(\mu'',\mu_0)
\right] \nn\\
&+& 
\left[1+ \int^\mu_{\mu_0} d\ln {\mu'}^2
P\left(N,\mu'\right)
\right]
f_{{\rm QC}}(N,\ln\frac{Q\theta}{u \mu_0}) \nn\\
&-& 
\frac{\alpha_s(\mu_0)}{2\pi}
\int^{\mu}_{\mu_0} d\ln {\mu'}^2
\left[1+ \int^\mu_{\mu'} d\ln {\mu''}^2
P\left(N,\mu''\right) \right]\,
\tilde{P}\left(N,\mu'\right) 
D(\mu',\mu_0)[2P^{Q}(N)]f(N,\mu_0)
\eea 
where in the last line, we have switched the order of the integrations, using 
\bea 
\int_{\mu_0}^\mu d \mu'  A(\mu') \int_{\mu_0}^{\mu'} d \mu'' B(\mu'') 
= \int_{\mu_0}^{\mu} d \mu'' B(\mu'')  \int_{\mu''}^\mu d \mu'  A(\mu')  \,.
\eea 

Iterate the procedure, we will arrive at  
\bea 
&& D(\mu,\mu_0) f_{\rm QC}(N,\ln \frac{Q\theta}{u \mu_0} ) 
+R(\mu,\mu_0) \nn \\ 
&-& \lim_{n\to \infty} 
\int_{\mu_{n-1}}^\mu d\ln\mu_n^2 P(N,\mu_n) \dots \int^{\mu}_{\mu1} d\ln\mu_2^2 P(N,\mu_2)
\left[D(\mu_1,\mu_0) f_{\rm QC}(N,\ln \frac{Q\theta}{u \mu_0} ) 
+R(\mu_1,\mu_0) \right]
\nn \\
&=& {\cal D}(\mu,\mu_0)f_{\rm QC}(N,\ln\frac{Q\theta}{u\mu_0})
- \frac{\alpha_s(\mu_0)}{2\pi} 
\int_{\mu_0}^\mu d\ln\mu'^2
{\cal D}(\mu,\mu')\tilde{P}(N,\mu')D(\mu',\mu_0)[2P^{Q}(N,\mu_0)]f(N,\mu_0) \,, \nn \\ 
\eea 
where ${\cal D} = \exp\left[\int_{\mu_0}^\mu d\ln\mu'^2 P(N,\mu') \right]$ is defined in Eq.~(\ref{eq:D}). 
We note that 
\bea 
&&
\lim_{n\to \infty} \frac{1}{n!}\left({ {\rm min}_{\mu} P(N,\mu)}\right)^{n-1} f_{\rm QC}(N,\ln\frac{Q\theta}{u\mu_0}) \to 0  
\nn \\ 
&& 
<
\lim_{n\to \infty} 
\int_{\mu_{n-1}}^\mu d\ln\mu_n^2 P(N,\mu_n) \dots \int^{\mu}_{\mu1} d\ln\mu_2^2 P(N,\mu_2)
\left[D(\mu_1,\mu_0) f_{\rm QC}(N,\ln \frac{Q\theta}{u \mu_0} ) 
+R(\mu_1,\mu_0) \right] \nn \\
&&
< 
\lim_{n\to \infty} \frac{1}{n!}\left({ {\rm max}_{\mu} P(N,\mu)}\right)^{n-1} f_{\rm QC}(N,\ln\frac{Q\theta}{u\mu_0}) \to 0 \,. 
\eea 
Here we have assumed that the moment of the PDF is bounded and thus the limit vanishes as $n\to \infty$. 

Therefore, we conclude that 
\bea 
&&f_{\rm QC}(N,\ln\frac{Q\theta}{u\mu})= D(\mu,\mu_0) f_{\rm QC}(N,\ln \frac{Q\theta}{u \mu_0} ) 
+R(\mu,\mu_0) 
\nn \\ 
&=& {\cal D}(\mu,\mu_0)f_{\rm QC}(N,\ln\frac{Q\theta}{u\mu_0})
- \frac{\alpha_s(\mu_0)}{2\pi} 
\int_{\mu_0}^\mu d\ln\mu'^2
{\cal D}(\mu,\mu')\tilde{P}(N,\mu')D(\mu',\mu_0)[2P^{Q}(N,\mu_0)]f(N,\mu_0) \,. \nn \\ 
\eea

Since $f_{\rm QC}(N,\ln\frac{Q\theta}{u\mu_0})$ and $R$ are independent and the solution should hold for arbitrary constant in $f_{\rm QC}(N,\ln\frac{Q\theta}{u\mu_0})$, then we could identify 
\bea
&&D = {\cal D} = \exp\left[\int_{\mu_0}^\mu d\ln\mu'^2 P(N,\mu') \right]
\,, \nn \\
&& R = - \frac{\alpha_s(\mu_0)}{2\pi}
\int_{\mu_0}^\mu d\ln\mu'^2
 {\cal D}(\mu,\mu')\tilde{P}(N,\mu'){\cal D}(\mu',\mu_0)[2P^Q(N,\mu_0)]f(N,\mu_0) \,.  
\eea

The derivation is applicable to higher logarithmic accuracy by suitably adjusting the relation in the initial condition of Eq.~(\ref{eq:lnu}) at higher $\alpha_s$ orders. 
\end{widetext} 

\bibliographystyle{h-physrev}   
\bibliography{refs}

\end{document}